\documentclass[5p,times]{elsarticle}

\usepackage[utf8]{inputenc}
\usepackage[T1]{fontenc}
\usepackage{amssymb,amsmath,array}  % amsthm
\usepackage[tight,footnotesize]{subfigure}
\usepackage{graphicx}      % include this line if your document contains figures
\graphicspath{{figures/}}

\newcommand{\rca}{$RC.1$}
\newcommand{\rcb}{$RC.2_{clu}$}

\newcommand{\rcai}{$RC.1.inp$}
\newcommand{\rcbi}{$RC.2_{clu}.inp$}

\begin{document}

\begin{frontmatter}

\title{Reservoir Computing for Detection of Steady State in Performance Tests of Compressors}

%\title{Elsevier \LaTeX\ template\tnoteref{mytitlenote}}

\author{Eric Aislan Antonelo \corref{mycorrespondingauthor}
	 %\fnref{myfootnote}
} 
\address{Interdisciplinary Centre for Security, Reliability and Trust (SnT), University of Luxembourg, Luxembourg}

\ead{ericaislan.antonelo@uni.lu}

\author{Carlos Alberto Flesch}
\address{LabMetro, Department of Mechanical Engineering, UFSC, Brazil}

\author{Filipe Schmitz}
\address{Whirlpool S.A., EMBRACO, Joinville, SC, Brazil}

\begin{abstract}
Fabrication of devices in industrial plants often includes undergoing quality assurance tests or tests that seek to determine some attributes or capacities of the device. 
For instance, in testing refrigeration compressors, we want to find the true refrigeration capacity of the compressor being tested. Such test (also called an episode) may take up to four hours, being an actual hindrance to applying it to the total number of compressors produced. 
This work seeks to reduce the time spent on such industrial trials by employing Recurrent Neural Networks (RNNs) as dynamical models for detecting when a test is entering the so-called steady-state region. Specifically, we use Reservoir Computing (RC) networks which simplify the learning of RNNs by speeding up training time and showing convergence to a global optimum.
Also, this work proposes a self-organized subspace projection method for RC networks which uses information from the beginning of the episode to define a cluster to which the episode belongs to. This assigned cluster defines a particular binary input that shifts the operating point of the reservoir to a subspace of trajectories for the duration of the episode. This new method is shown to turn the RC model robust in performance with respect to varying combination of reservoir parameters, such as spectral radius and leak rate, when compared to a standard RC network.

\end{abstract}


\end{frontmatter}

% \linenumbers

\section{Introduction}

A performance test of a refrigeration compressor, for instance, to find out the cooling capacity of the device, has an important role in the research and development of methods to achieve increasingly high levels of energy efficiency in the context of refrigeration thermal machines. 
The global market indeed requires the continuous enhancement of these compressors, which can be confirmed by the fact that the current leader in compressor production has halved the energy consumption requirements over the last two decades \cite{Coral2015}.
Thus, the compressor performance test is an essential procedure in the advancement of these technologies. Besides, it also ensures that efficiency settings accorded through contracts are respected: the refrigeration (cooling) capacity is one of the main parameters obtained during a performance test, and is also very important for client companies that buy compressors to build thermal machines.

There are different methods to obtain the refrigeration capacity of a compressor. The ISO 917 standard \cite{iso917} requires that the measurement of the refrigeration capacity should be done under steady-state conditions: the measured quantities should be within a predefined margin for a minimum interval of one hour \cite{iso917}. In practice, however, the complete test duration is two and a half hours on average, and can take up to four and a half hours in some cases. In addition, the production volume is very high in a single plant (in the order of tens of thousands), making  these types of standardized tests impracticable to be implemented for the whole set of compressors produced, but only to a small sample of it \cite{Coral2015}.

In this context, it is very desirable to employ techniques that can detect when the refrigeration capacity signal, obtained through measurements during the performance test in the compressor, enters the steady state region. This has the potential to reduce the time needed to run these performance tests, which in turn increase the productivity of the plant and/or the number of compressor samples to be performance-tested.

The objective of this work is to design a dynamic classification model that can be used to detect this steady state entrance. 
Other important available measurement signals are used as input to the model (as in \cite{Penz2015}) in addition to the cooling capacity: the compressor shell temperature and the compressor suction pressure. 
Considering only the current input measurements is not enough to create such model. That is why previous approaches \cite{Penz2015} have used black-box models such as feedforward neural networks with extra (specially handcrafted) input features made of moving averages and derivatives of the original signals, totaling 16 input dimensions to the model. In contrast to this, this work employs Recurrent Neural Networks (RNNs) \cite{schmidhuber2015deep,Maass2002a,Jaeger2001a,hochreiter1997LSTM} to naturally cope with the dynamical intricacies of the task without the need to create special input features.

\begin{figure}
 \centering  
 \includegraphics[width=5.8cm]{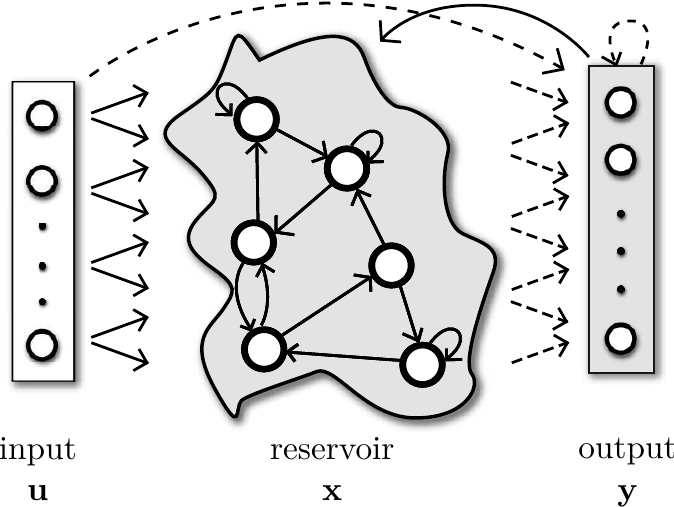}  % 6.1cm
 \caption{Reservoir Computing (RC) network. The reservoir is a non-linear dynamical system usually composed of recurrent sigmoid units. 
 Solid lines represent fixed, randomly generated connections, while dashed lines represent trainable or adaptive weights. 
}
 \label{fig:rc_network}
 \label{fig:reservoirPic}
\end{figure}

RNNs can provide a type of state-dependent computation much like cortical functioning in the brain \cite{Buonomano2009}, where the trajectory of a high-dimensional dynamical system reflects both the current input as well as previously received input stimuli.
Reservoir Computing (RC) \citep{Verstraeten2006a} is a term recently coined to designate this paradigm of computation based on transients of a fixed dynamical system (such as an RNN).
Echo State Networks (ESNs) \cite{Jaeger2004} and networks based on backpropagation-decorrelation learning \cite{Steil2006} were the first RC models proposed using analog neurons, while Liquid State Machines (LSMs) \cite{Maass2002a} basically consist of dynamical reservoirs made of spiking neurons. 
In RC, the network (see Fig. \ref{fig:rc_network}) is composed of a recurrent high-dimensional pool of neurons, with randomly generated and fixed synaptic weights, called reservoir, and a linear adaptive readout output layer which projects the reservoir states to the actual system's output. As only the output layer needs to be trained, usually via linear regression methods, the training is simplified and global convergence guaranteed. On the other hand, traditional methods to train RNNs, such as BPTT (backpropagation through time \citep{Werbos1990}), have slow training and no global convergence guarantee. 

The reservoir can be viewed as a dynamic nonlinear kernel that projects the input to a high-dimensional dynamic space, where linear regression or classification is usually enough for various tasks. Many applications relying on the powerful temporal processing capabilities of RC exist: navigation and localization of mobile robots in partially observable environments \citep{Antonelo2014}, 
% Antonelo2011
periodic signal generation with nanophotonic reservoir computing \citep{Fiers2014}, hierarchical control of robotic arms \citep{Waegeman2013}, speech recognition \citep{Triefenbach2013}, modeling of soft-sensors for offshore oil production platforms \cite{Antonelo2016}, etc.

This work also proposes a new RC architecture which uses a priori knowledge to constrain the dynamical reservoir space to a predefined subspace. This subspace is defined by binary inputs to the RC model as in \cite{Antonelo2014} for the task of robot navigation. In our proposal, the a priori knowledge will be given by the application of an unsupervised learning mechanism such as k-means clustering on an initial period (first 13 minutes) of the performance test. We will see that the resulting model is more robust with respect to the parameters of the RC network (spectral radius and leak rate - see Section~\ref{sec:methods}), producing good generalization performance with less dependence on these parameters when compared to the standard RC network.

This paper is organized as follows: Section~\ref{sec:compressors} explains the performance test process in refrigeration compressors.
Next, related works in the literature are compared against the current approach (and its novel aspects) in Section~\ref{sec:related_work}.
Section~\ref{sec:methods} presents the reservoir computing model and the proposed self-organized subspace projection method.
The results are shown and analysed in Section~\ref{sec:experiments} .
Section~\ref{sec:conclusion} concludes this work and gives future research directions.

\section{Performance tests in Refrigeration Compressors}\label{sec:compressors}
The performance of a refrigeration compressor is measured on a specific rig that simulates a refrigeration system with several measured and controlled variables.
The cooling (or refrigeration) capacity is an indirect measurement defined by the product of the mass flow rate of the refrigerant fluid in the compressor and the enthalpy difference between two specific points in the refrigeration circuit \cite{iso917}. The enthalpy values, in turn, are constants defined by the operating condition of the compressor and the refrigerant fluid. The measurement of the mass flow rate can be done by nine different methods \cite{iso917}. One of them is obtained by measuring the mass flow rate directly through a commercial mass flow meter that reaches the minimum measurement uncertainty defined by international standards. Another commonly used method for mass flow measurement (which is independent of the former method) is based on heat balance using a calorimeter \cite{iso917}.

\begin{figure}
 \centering  
 \includegraphics[width=7.8cm]{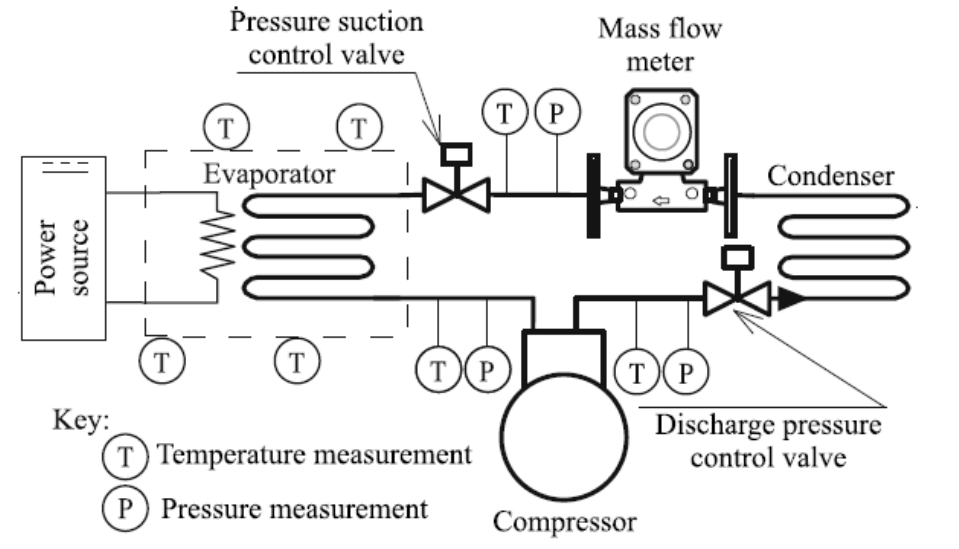}  % 6.1cm
 \caption{
    Simplified schematic of the test rig.
 }
 \label{fig:compressor}
\end{figure}

A simplified schematic of a test rig to measure a compressor performance (initially described by \cite{Flesch2010}) is shown in Figure \ref{fig:compressor}.
Basically a rig contains a compressor under test, suction and discharge pressure controller (that can be done through valves), condenser, mass flow meter and a calorimeter (represented by the area inside the dashed rectangle) and pressure and temperature transducers installed at different parts of the circuit to monitor test conditions.

Due to the complexity of the system to reach the steady state condition, the performance test takes 2.5 hours on average. 
The most critical variables that affect the final result of the performance test are the compressor shell temperature, the suction pressure and the discharge pressure. 
Both the suction pressure and the discharge pressure reach the steady state when their measurements are kept inside the interval defined by $ \pm 1\%$ of the set point (that is a value chosen depending on the compressor operation condition). 
Due to the high sensibility of the compressor to those variables, if their values reach outside the above interval, the steady state inference is restarted to avoid unreliable results. 

As each compressor has a different steady state value for the shell temperature, there is no set point to be defined. The shell temperature is assumed to be in steady state when its value is inside the interval defined by $ \pm  1^{\circ}C $ of the final (unknown) value. For instance, if the final value is to be 63 $^{\circ} C$, the steady state takes place in $[62 ^{\circ}C, 64
 ^{\circ}C]$.
The same idea applies to the cooling capacity, which has no set point. Its steady state takes place when its current measurement is inside the $ \pm  2\% $ margin of unknown \emph{final cooling capacity}, defined as the average of the signal over the last 45 minutes of the test.
 See Section~\ref{sec:dataset} for more details on the dataset generation.

\section{Related work and Contribution} \label{sec:related_work}
In \cite{Penz2012}, a hybrid Fuzzy-Bayesian network approach is used for predicting the final (unknown) refrigeration capacity of the compressor based on a single input corresponding to the current refrigeration capacity. The resulting system can also be used to reduce the time of a compressor performance test. 
Their approach makes assumptions on the behavior of the input signal, yielding a restricted applicability in this sense. They also only show the application to a dataset of very limited size (for instance, the test dataset contains no more than 30 performance trials) with trials executed on only one compressor model.

The work presented in \cite{Coral2015} also aims at inferring the final value of the refrigeration capacity, but it uses a completely different measurement independent of the refrigeration test rig from Fig.~\ref{fig:compressor}: it is based on the pressure increase rate within a vessel of known volume measured when the compressor is turned on. This value is shown to be correlated to the true refrigeration capacity of the compressor. Their method allows this inference is less than 7 seconds, since it does not use the conventional test rig and the refrigerant fluid. Their predictive model is composed of a committee of feed-forward neural networks having 8 inputs: the pressure increase rate and compressor design parameters; and it is compressor model-dependent, i.e.,  
different committees have to be trained for each compressor model.

In later work, Penz \cite{Penz2015} proposes a method composed of a committee of feed-forward networks trained by backpropagation for predicting the moment of steady state entrance during a performance test (similarly to this work). The method also uses the test rig of Fig.~\ref{fig:compressor} (unlike \cite{Coral2015}).
Furthermore, it has 16 inputs made of moving averages, derivatives and standard deviation of three variables: cooling capacity, shell temperature and suction pressure. These handcrafted input features were created in order to account for the static (memoryless) nature of the multilayer perceptron networks.
The committee method used is suitable when considering small training sets (only 40 performance tests in \cite{Penz2015}) in order to avoid overfitting. On the other hand, their method, based on backpropagation, can not easily handle large training sets in practice due to the considerably long training process.
In addition, \cite{Penz2015} uses different models of compressors in the training set, yielding signals of different behaviors and magnitudes. Their preprocessing includes normalization of each input variable according to a priori knowledge of the compressor model being tested. The resulting normalized signal (of the cooling capacity) converges to a value around $1$ at the end of the performance test in a way independent of the compressor model, facilitating the learning task of the network.
In practice, the requirement of a priori knowledge (in the form of some parameters of the compressor model for signal normalization) is not desired, since the compressor performance test itself is devised for such end: uncovering these unknown parameters (e.g. cooling capacity) of the compressor being tested.

In contrast to previous approaches, the current work innovation is four-fold:
\begin{itemize}
	\item combines self-organization of k-means clustering and Reservoir Computing networks to effectively and quickly learn the detection (classification) task. The approach of subspace projection given by k-means and RC together is novel as far as the authors know, forming a general black-box method applicable to many other time series-based detection problems.
	
	\item is totally data-driven. In the context of steady-state detection, the dataset originates from many different models of compressors (see Fig.~\ref{fig:hist_refcap}, for instance) - around 70 models, significantly more than related previous work; and no a priori knowledge from the compressor specification is used to normalize the signal (unlike in \cite{Penz2015}). Only a single RC network is employed for all 70 compressor models.
	
	\item it is trained on a much larger heterogeneous dataset of 614 compressor performance tests when compared to previous works, in a quicker and more effective way (due to the global convergence properties of the RC training approach);

	\item no lagged inputs, moving averages, compressor design parameters or any hand-crafted features are necessary using the dynamical RC approach, but only the current input measurements are used due to the inherent short-term memory present in reservoirs.

	\end{itemize}

\section{Methods}\label{sec:methods}

%%%%%%%%%%%%%%%%%%%%%%%%%%%%%%%%%%%%%%%%%%%%%%%%%%%%%%%%%%%%%%%%%%%%%%%%%%%%%%%%
\subsection{Reservoir Computing} \label{sec:reservoir}
The RC model we use is based on the Echo State Network (ESN) approach \cite{Jaeger2001a}. The state update equation for the reservoir is given by:
\begin{equation}\label{eq:stateUpdate}  
\mathbf{x}(t+1) = f((1-\alpha)\mathbf{x}(t) + \alpha(\mathbf{W}_{\mathrm{in}}  \mathbf{u}(t) +
\mathbf{W}_{\mathrm{res}}  \mathbf{x}(t) + \mathbf{W}_{\mathrm{bias}}^{\mathrm{res}} ) ),
\end{equation}
where: $\mathbf{u}(t)$ represents the input at time $t$; $\mathbf{x}(t)$ is the reservoir state; $\alpha \in (0,1] $ is the leak rate%\cite{Schrauwen2007b}
; and $f()=\tanh()$ is the hyperbolic tangent activation function; $\mathbf{W}_{\mathrm{in}}$ and $\mathbf{W}_{\mathrm{bias}}^{\mathrm{res}}$ are the weight matrices from input and bias to reservoir, respectively and $\mathbf{W}_{\mathrm{res}}$ represents the recurrent connections between internal nodes of the reservoir. The initial state is $\mathbf{x}(0) = \mathbf{0}$. A standard reservoir equation (without the leak rate) is found when $\alpha = 1$.

The output of the RC network $\mathbf{y}(t)$ is given by a linear combination of the reservoir states plus a bias, which can be post-processed by a function $g$:
\begin{equation}\label{eq:output}  
 \mathbf{y}(t+1)  =  g(\mathbf{W}_{\mathrm{out}}\mathbf{x}(t+1)+\mathbf{W}_{\mathrm{bias}}^{\mathrm{out}}).
\end{equation}

The non-trainable weights $\mathbf{W}_{\mathrm{in}}$, $\mathbf{W}_{\mathrm{res}}$ and $\mathbf{W}_{\mathrm{bias}}^{\mathrm{res}}$ are randomly generated from a Gaussian distribution $N(0,1)$ or a uniform discrete set $\{-1,0,1\}$. 
After this random initialization, the matrix $ \mathbf{W}_{\mathrm{in}}^{\mathrm{res}} $ ($\mathbf{W}_{\mathrm{bias}}^{\mathrm{res}}$ ) is scaled by the parameter called input scaling $\upsilon_{\mathrm{inp}}$ (bias scaling $\upsilon_{\mathrm{bias}}$). 
Additionally, the $\mathbf{W}_{\mathrm{res}}$ matrix is rescaled so that the reservoir has the echo state property \cite{Jaeger2001a}, that is, the spectral radius
$ \rho( \mathbf{W}_{\mathrm{res}}) $ (the largest absolute eigenvalue) of the linearized system is smaller than one \cite{Jaeger2001a}. This means that the reservoir should have a fading memory such that if all inputs are zero, the reservoir states also approach zero within some time period.
The leak rate of the reservoir controls the timescale of its neurons $\mathbf{x}(t)$ \cite{Jaeger2007}, i.e., how slowly or how fast they react to the incoming input stream $\mathbf{u}(t)$. Slow reservoirs, with more memory, are obtained when $\alpha$ approaches zero and the \emph{quickest} reservoir is given by $\alpha = 1$. Resampling the input or adjusting the leak rate can help matching the timescale of the reservoir to the input timescale or the task (output) timescale.
The configuration of the reservoir parameters are given in Section~\ref{sec:experiments}.

Next, consider the following notation: $n_i$ is the number of inputs; $n_r$ is the number of neurons in the reservoir; $n_o$ is the number of outputs; $n_s$ is the number of samples.
The training of the output layer is done by using the \textbf{Ridge Regression} method \cite{Bishop2006}, also called \emph{Regularized Linear Least Squares} or \emph{Tikhonov regularization} \cite{Tychonoff1977}: 	
\begin{eqnarray} \label{eq:ridge_regression} 
  \mathbf{{W}}_{\mathrm{out}} = (\mathbf{X^{\top}}\mathbf{X} + \lambda\mathbf{I})^{-1}\mathbf{X^{\top}} \hat{\mathbf{Y}} 
\end{eqnarray}
%%%%%%%%%%%%%%%
where: $\mathbf{X}$ is the matrix of size $ n_s \times (n_r+1) $ with the generated reservoir states collected row-wise (using (\ref{eq:stateUpdate})) where the last column of $\mathbf{X}$ is composed of 1's (representing a bias); $\lambda$ is the regularization parameter.
The desired outputs, i.e., -1 (not in steady-state) or +1 (in steady-state), are collected row-wise into a matrix $ \hat{\mathbf{Y}}$.

% \cite{Bishop2006}. 
Note that the other matrices ($ \mathbf{W}_{\mathrm{res}}, \mathbf{W}_{\mathrm{in}}, \mathbf{W}_{\mathrm{bias}}^{\mathrm{.}}$) are not trained at all. The last two matrices (connections from input/bias to reservoir) are configured in Section~\ref{sec:experiments}.
The learning of the RC network is a fast process without local minima, which is not the case for algorithms such as BackPropagation-Through-Time (BPTT). 

%%%%%%%%%%%%%%%%%%%%%%%%%%%%%%%%%%%

\subsection{Self-organized Subspace Projection} \label{sec:subspace}

\begin{figure}
 \centering  
 \includegraphics[width=5.8cm]{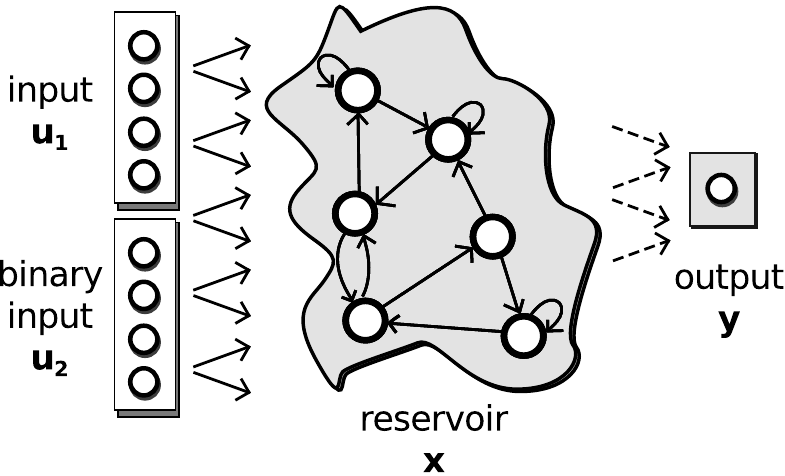}  % 6.1cm
 \caption{RC network with inputs divided in original inputs $u_1$ and binary inputs $u_2$. }
 \label{fig:rc_subspace}
\end{figure}

\begin{figure}
 \centering  
 \includegraphics[width=7.8cm]{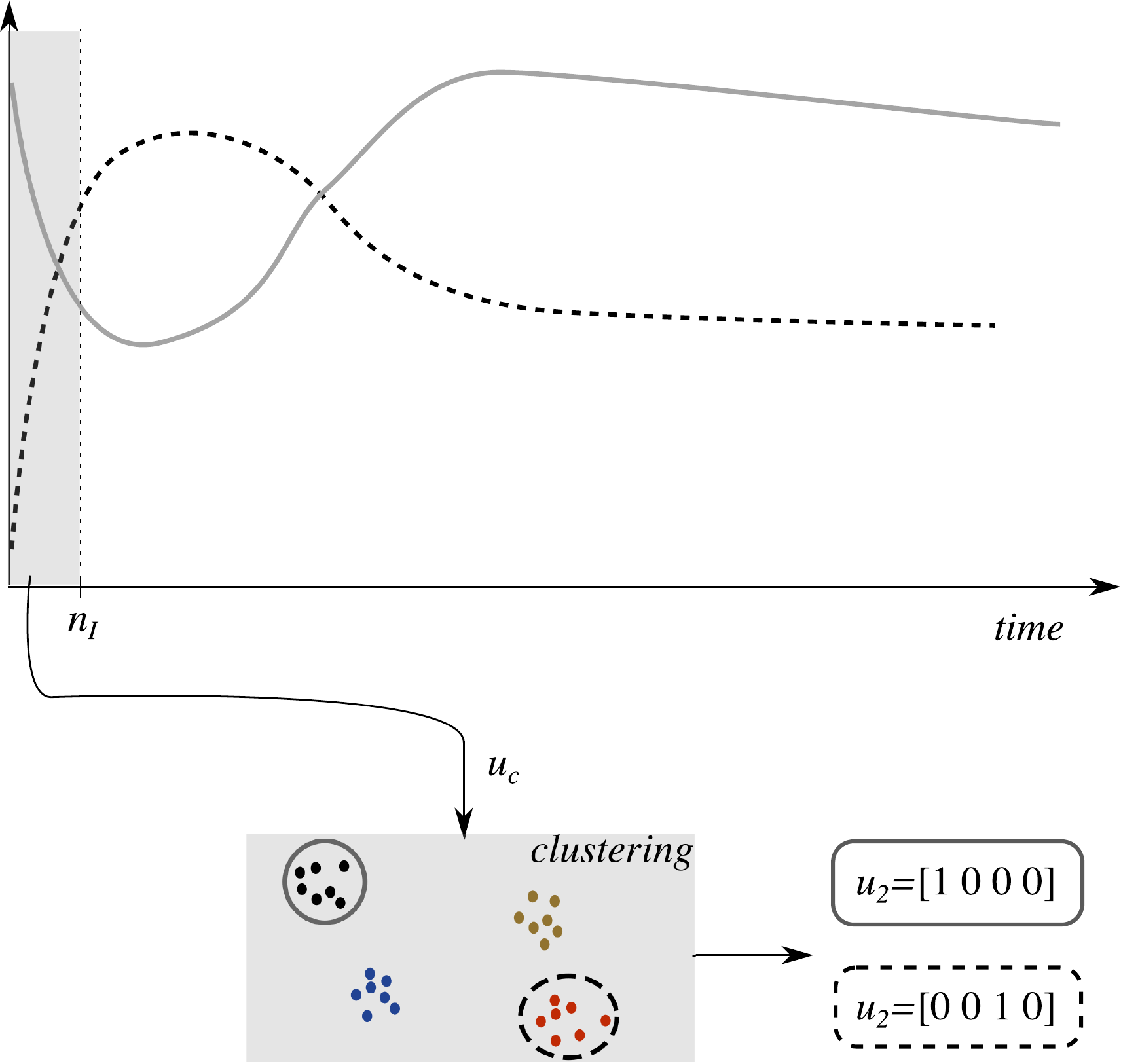}  % 6.1cm
 \caption{Generating a priori information $u_2$ for reservoir subspace projection through unsupervised learning in the very beginning of the episode.  
 The vertical axis in the top plot represents any input signal $u_1$ (in our case, the cooling capacity).
 $u_c$ is the $(n_I \times 1)$-dimensional input to the clustering model.}
 \label{fig:clustering}
\end{figure}

We also use an RC architecture, first proposed in \cite{Antonelo2011b} in the context of robot navigation, in order to introduce some sort of a priori knowledge into the reservoir operation. This architecture has binary input units whose sole function is to constrain the reservoir trajectory to predefined subspaces (Fig.~\ref{fig:rc_subspace}). For our particular application, there is only one output with the sign nonlinearity as nonlinear post-processing function $g$. 
Furthermore, the binary input pattern is found in a self-organized way by clustering the initial operation of the compressor test. 
This is done by applying the K-means clustering algorithm on an initial temporal slice (e.g., the first $n_I=13$ minutes) of the refrigeration capacity signal. With this, we can decide to which cluster the current performance test belongs to (see Fig.~\ref{fig:clustering}). 
For instance, if the initial behavior of the signal is assigned to cluster 1, from a total of 4 clusters, then $\mathbf{u_2}(t) = \left [ 1, 0, 0, 0 \right ]^T$ for $t=0,....,n_e$ where $n_e$ is the total number of samples in the performance test. Every performance test corresponds to an episode of variable size $n_e$, reseting the reservoir state to its initial value $\mathbf{x}(0) = \mathbf{0}$ before starting reservoir simulation for each episode.

\begin{figure}[t!]
	\centering  
	\includegraphics[width=7.3cm]{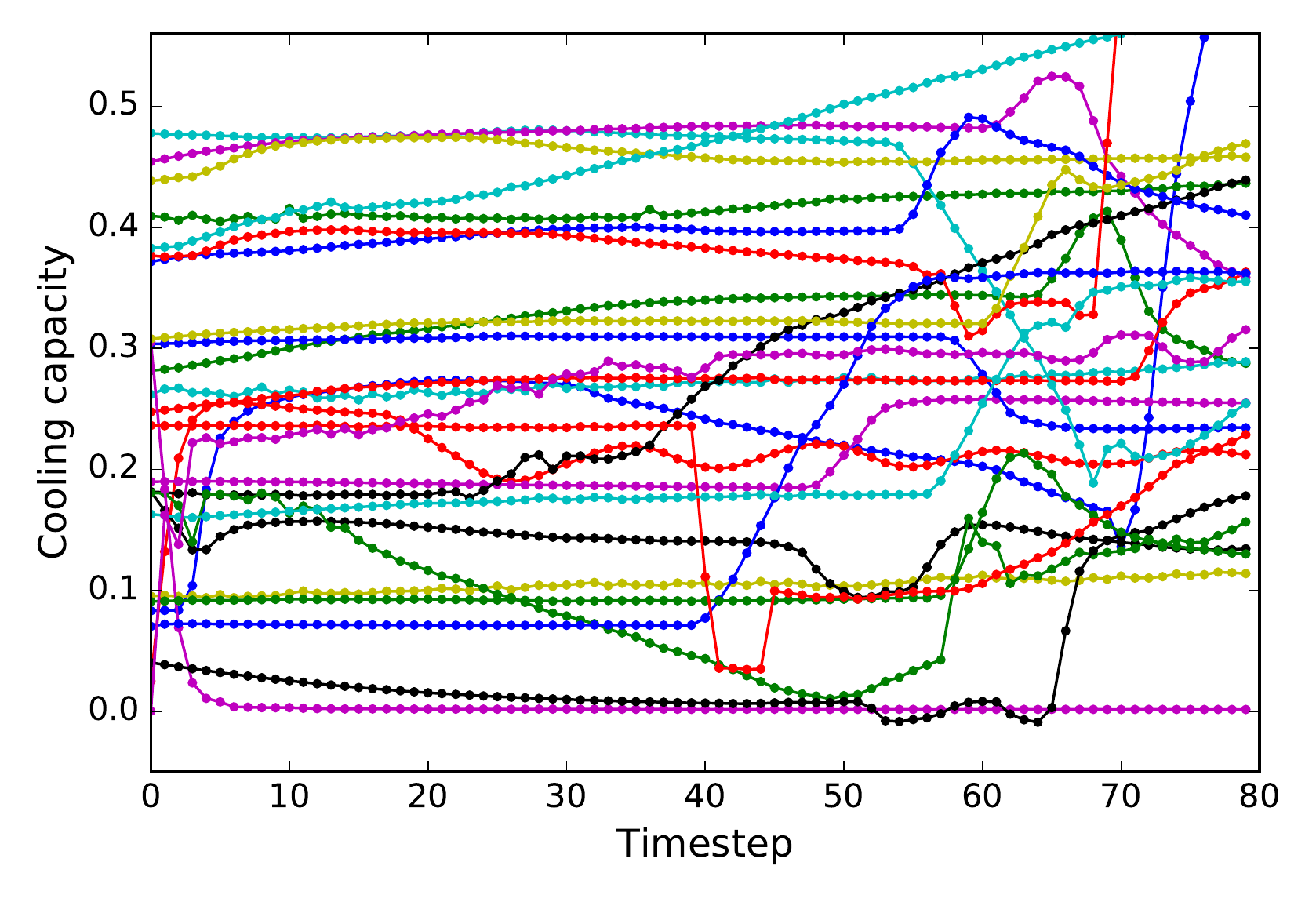}  % 6.1cm 
	\caption{General overview of the diversity of dynamical behaviors and range for the normalized cooling capacity during an
		initial time period corresponding to the first 80 samples (i.e., $n_I=13$ minutes).
Only 26 performance tests (from different compressor models) 
from the training dataset are shown. }
	\label{fig:initial_refcap}
\end{figure}

The unsupervised training (of the k-means clustering) uses all training samples, but only in the interval $t=0,...,n_I$. This initial period often contains a diverse set of dynamical behaviors originating from many different models of compressors and operating conditions
(see Fig.~\ref{fig:initial_refcap}),
which could give some a priori knowledge of the type of compressor is currently being tested. This also implies that no steady-state detection is possible before $t=n_I$ since these initial samples will always be reserved to the unsupervised learning process and corresponding cluster assignment.

%%%%%%%%%%%%%%%%%%%%%%%%%%%%%%%%%%%%%%%%%%%%%%%%%%%%%%%%%%%%%%%%%%%%
\subsection{Dataset generation}\label{sec:dataset}
We are interested in three types of measurements throughout a compressor performance test: the \textbf{refrigeration (or cooling) capacity} (in Watts), \textbf{the suction pressure}, and \textbf{the shell temperature}. Each compressor test is comprised of a variable number of $n_e$ samples collected every 10 seconds (each sample contains measurements of the three aforementioned variables). These data are normalized to be in the range $[0,1]$ (e.g., dividing by the maximum corresponding value in the training set) and inserted into a structure composed of episodes (each episode corresponds to one compressor test where the input is the 3-dimensional vector $\mathbf{u_1}(t), t=0,...,n_e$). 
The fourth optional input variable ($setpoint$) is a binary signal generated from the controlled suction pressure signal. Its value is zero (0) if the pressure value is not located in the reference setpoint region. Note that the value of the reference setpoint ($ref$) for the suction pressure is known beforehand for a particular compressor test. Correspondingly, its value is one (1) if the pressure value is inside the setpoint region: $pressure \in [ref * (1-0.01), ref * (1+0.01)]$. Here, the pressure is assumed to have reached the reference setpoint ($ref$) when it is inside the $\pm1\%$ tolerance margin from $ref$.  An example can be seen in the forth plot from left to right in Fig.~\ref{fig:ensaio1} (or Fig.~\ref{fig:ensaio2}), showing both the suction pressure and the corresponding binary reference setpoint indicator ($setpoint(t)$).

The desired outputs $ \hat{y} \in \{-1,1\}$ are generated such that they indicate that the refrigeration capacity $cap(t)$ has reached the region considered the steady-state of the signal. This area corresponds to the $\pm2\%$ margin from the final cooling capacity $cap_f$, which in turn is computed as the average over the last 45 minutes of the $cap(t)$ in the episode ($cap_f = 1/45\sum_{t=n_e-45}^{n_e}cap(t)$). Thus, $ \hat{y}(t) = 1$ if $cap(t) \in [cap_f * 0.98, cap_f * 1.02] $, otherwise $\hat{y}(t) = -1$. This computation of the desired output also makes sure that there is only one switch from zero (0) to one (1) in the target signal $ \hat{y} $, disregarding intervals in which $cap(t)$ is only temporarily in the $\pm2\%$ margin of $cap_f$ (pseudo-code not shown here). In addition, another condition for $ \hat{y}(t) = 1$ is that $setpoint(t) = 1$, that is, $cap(t)$ can be in steady-state only after the suction pressure has reached its reference setpoint.

Note that $cap_f$ is not known a priori (as the pressure setpoint $ref$ is), but is computed from the training samples. Thus, the target $ \hat{y}(t) $ signal has predictive information in it computed from samples.
 
\section{Experiments} \label{sec:experiments}

The dataset was obtained from Embraco, totalling $1,023$ compressor performance tests, that are equivalent to $862,323$ samples (one sample collected each 10 seconds), or 99 days worth of measurements. The training, validation and test\footnote{\emph{Test} dataset refers to the data not used in the training of the RC model while compressor \emph{test} refers to the trial process executed on a compressor during which measurements are taken, also called here an episode.} datasets were selected randomly from these 1,023 compressor tests (or episodes) in the following proportion: 60\%, 20\% and 20\% respectively. 
\begin{figure}[t!]
	\centering  
	\subfigure[] {
		\includegraphics[width=7.3cm]{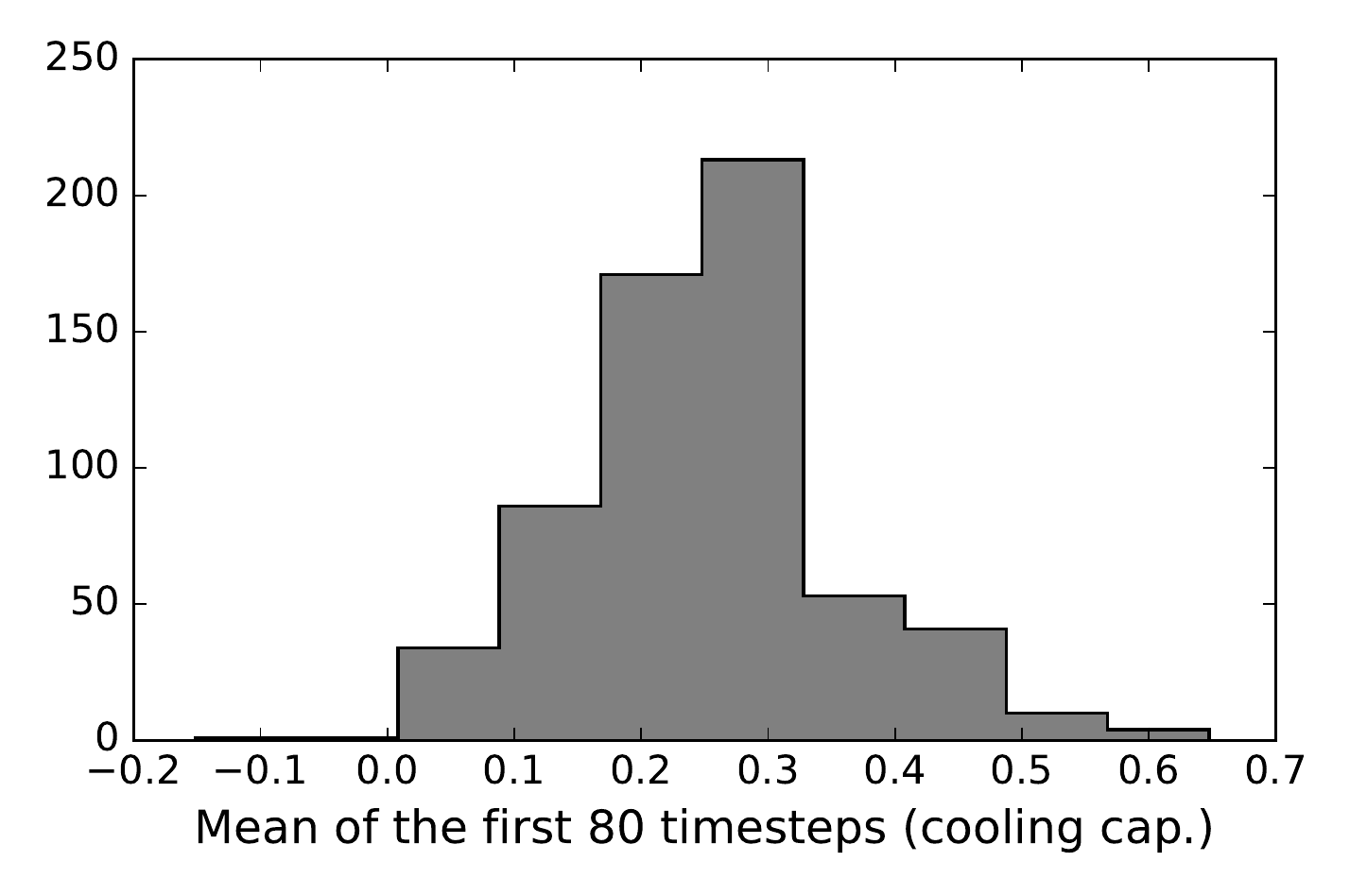} 
	}
	\subfigure[]{
		\includegraphics[width=7.3cm]{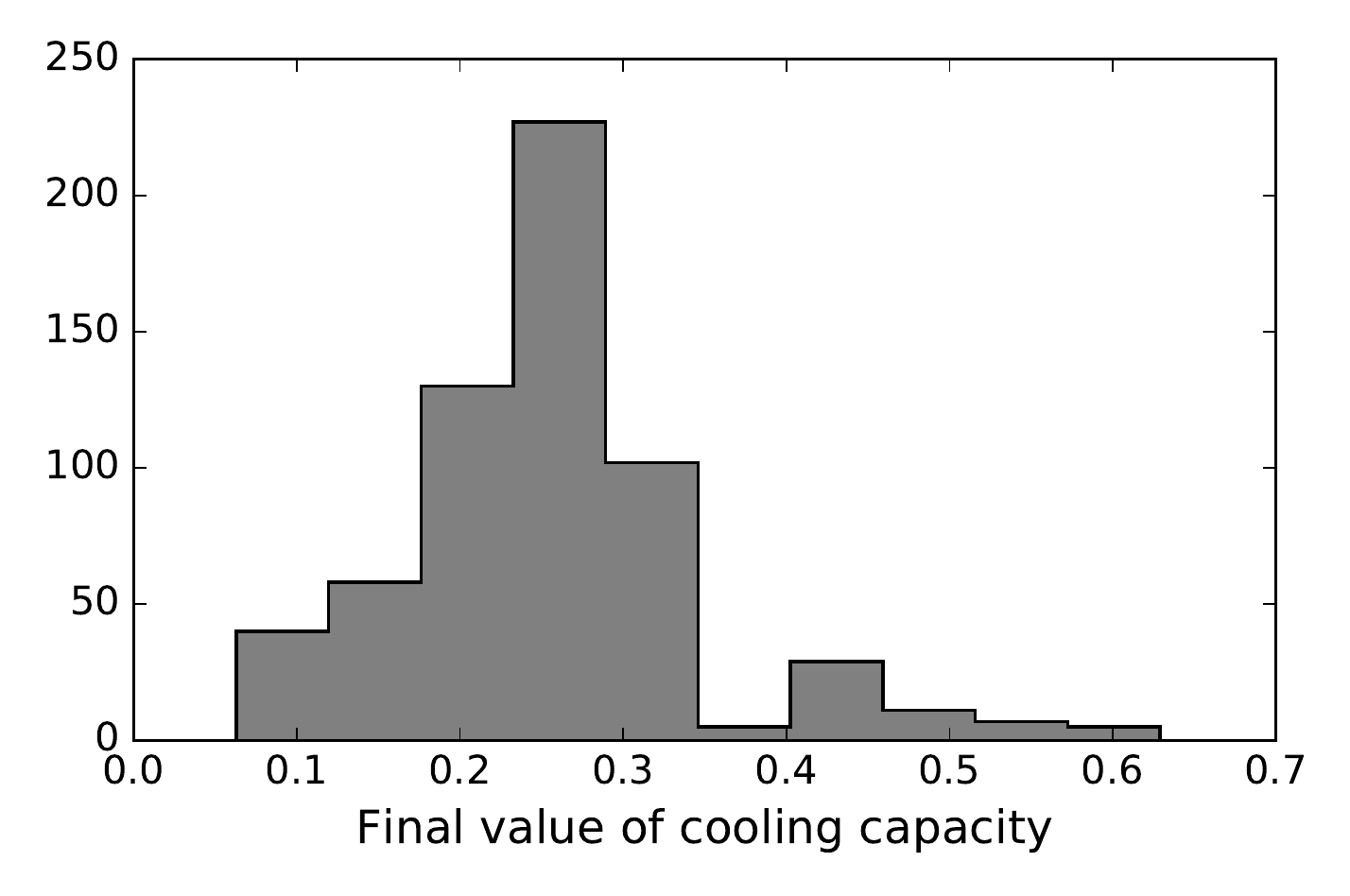}
	}
	
	\caption{Different compressor models with different initial and final values of the cooling capacity. (a) The distribution of the means of the cooling capacity for the initial 80 samples (13 minutes) in the training dataset (614 performance tests). (b) The distribution of the final values of the cooling capacity (when the performance test finishes). }
	\label{fig:hist_refcap}
\end{figure}
Besides, these data are originated from different types of compressors (around 70 types) and no a priori knowledge from the compressor specification is used to normalize the signal (unlike in \cite{Penz2015}). The only normalization used in the current work is given by a conventional division of an input variable by the maximum value that it takes in the training dataset. For instance, the diversity of the operating regimes of the different models of compressors can be seen in 
Fig.~\ref{fig:hist_refcap} that shows
 the histograms of the mean of the (normalized) refrigeration capacity during the first 13 minutes and of the final value for all compressor tests in the training dataset.

%%%%%%%%%%%%%%%%%%%%%%%%%%%%%%%%%%%%
\begin{figure*}[!htb]
 \centering  
\subfigure[] {
 \includegraphics[width=11.3cm]{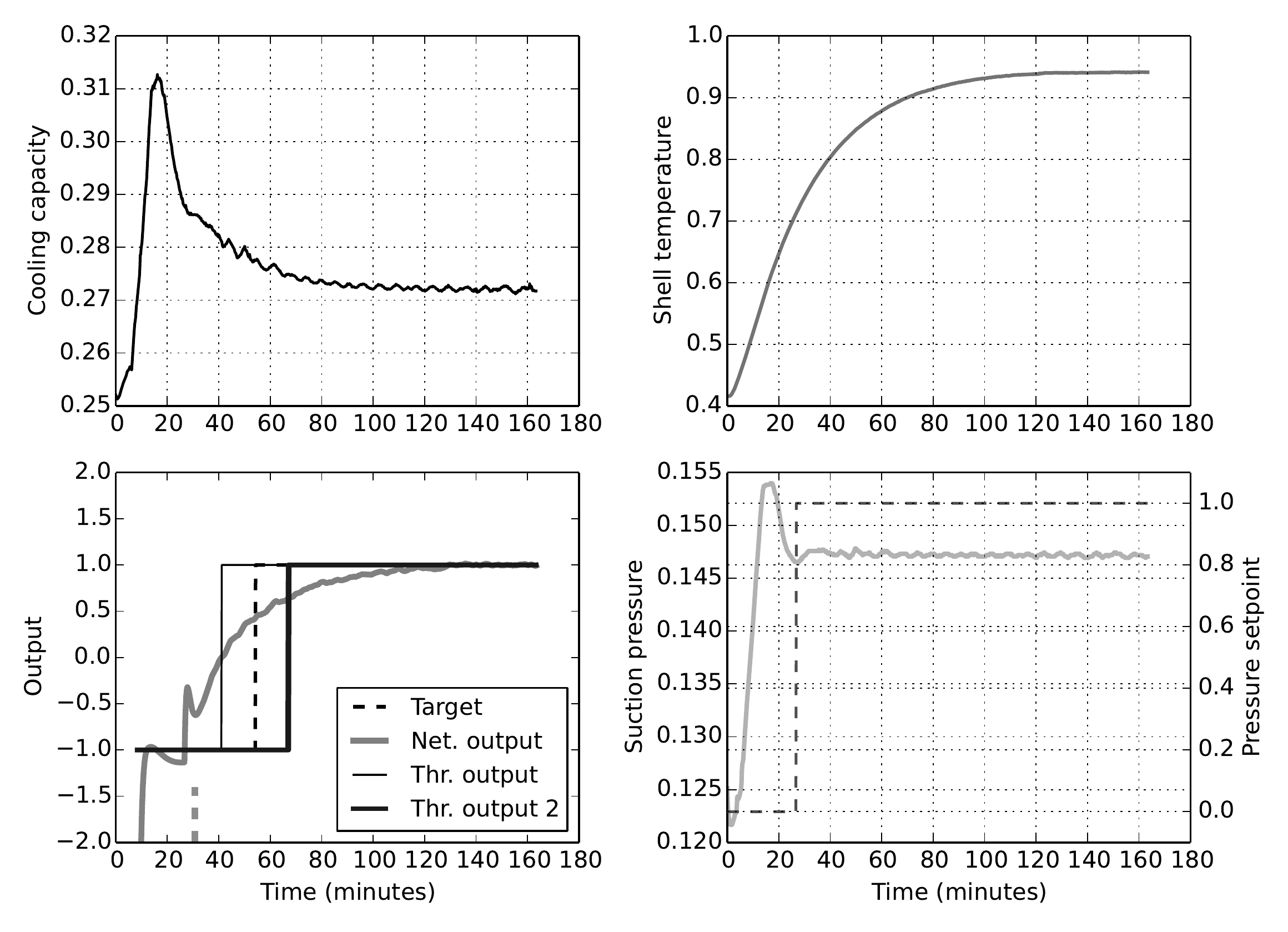}  % 6.1cm
 \label{fig:ensaio1}
}
\subfigure[] {
 \includegraphics[width=11.3cm]{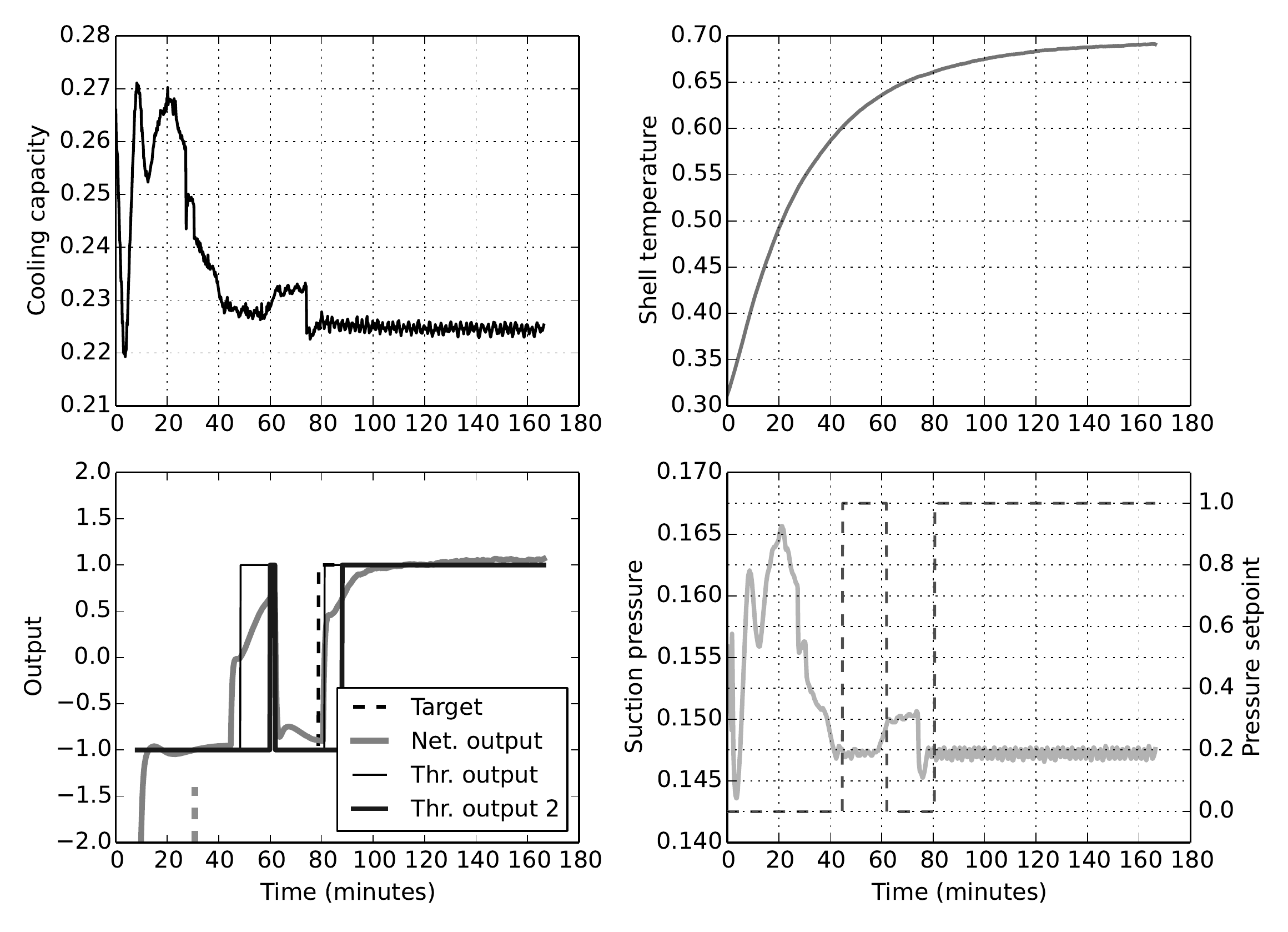}  % 6.1cm
 \label{fig:ensaio2}
}
 \caption{Input and output variables for two compressor performance tests. The plots with a grid show the 4 inputs to the RC model. The target output is given by a dashed line in the let plot of the second row together with the predicted outputs: the analog network output (thick grey line), thresholded output (grey line), thresholded output 2 (thick black line). For more information, see text.
 }
 \label{fig:ensaios}
\end{figure*}  
%%%%%%%%%%%%%%%%%%%%%%%%%%%%%%%%%%%%

\begin{figure} [htb!]
 \centering  
 \subfigure[\rcai] {
 \includegraphics[width=5.7cm]{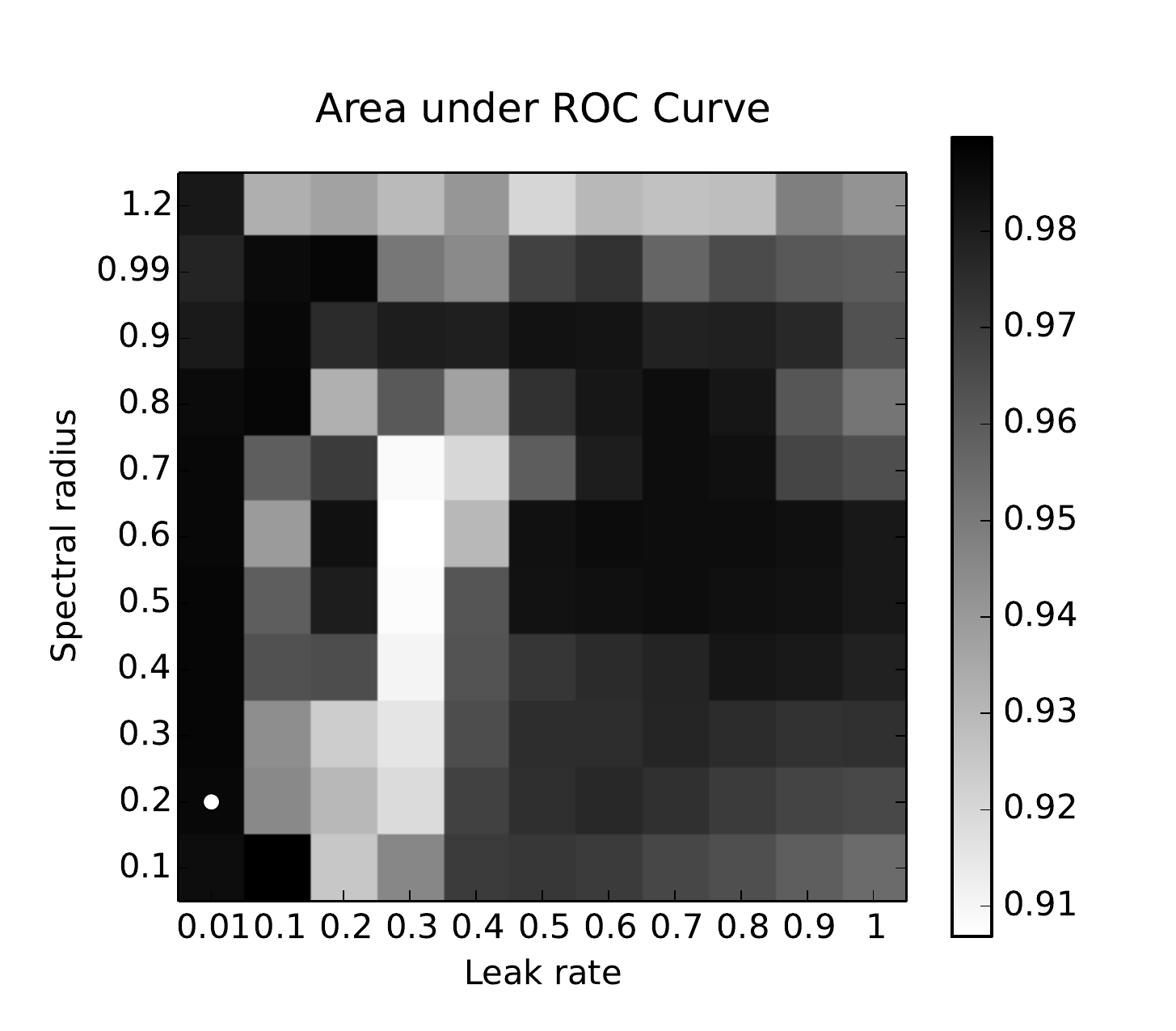}  % 6.1cm
 }
 \subfigure[\rcbi] {
 \includegraphics[width=5.7cm]{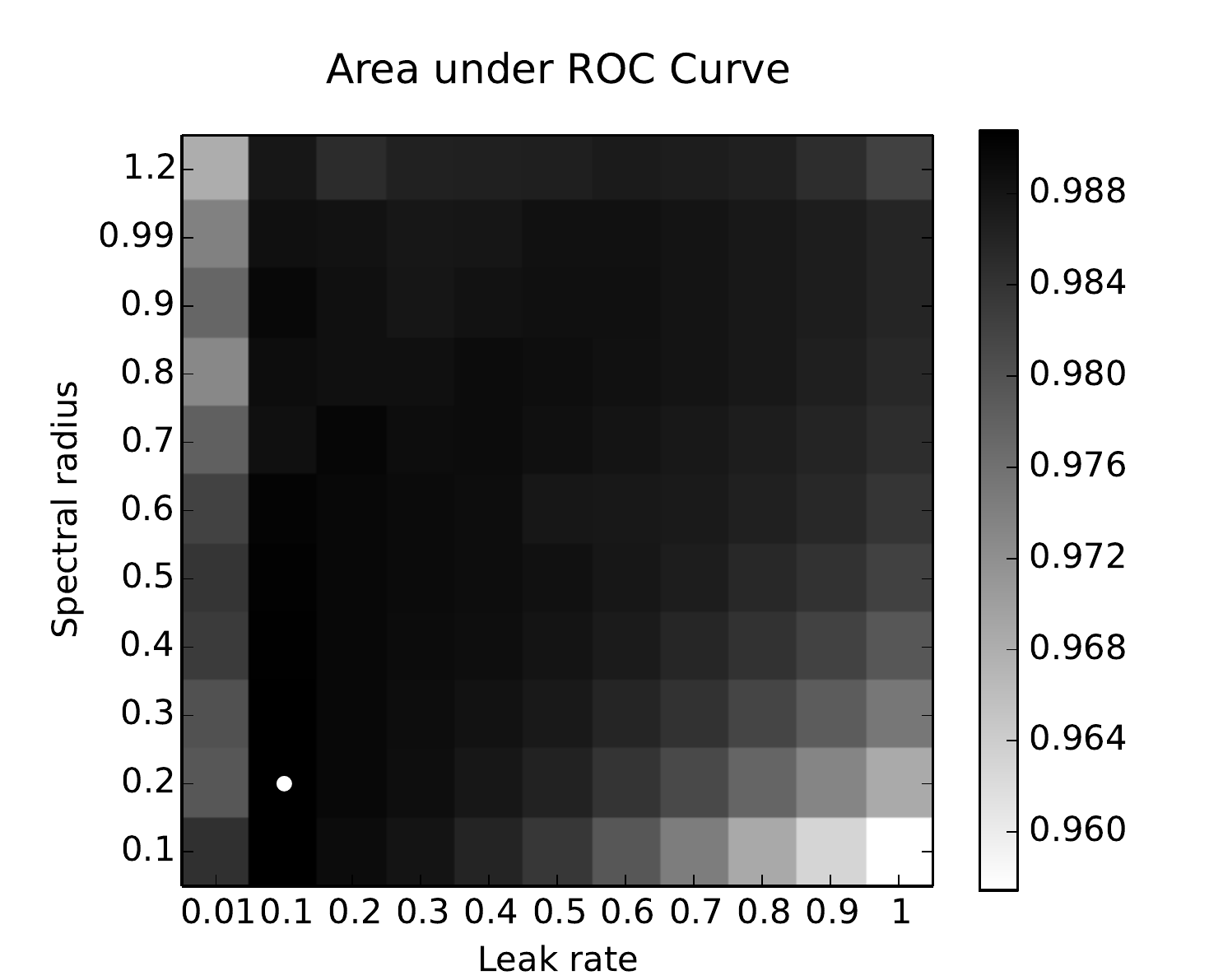}  % 6.1cm
 }
 \caption{Area under ROC for different settings of spectral radius and leak rate. The suction pressure setpoint indication is also used as input. The white dot marks the optimal parameter configuration.
 }
 \label{fig:auc_params}
\end{figure}

The RC model is trained on 614 episodes with the following settings: 600 reservoir units, leak rate $\alpha=0.1$, input scaling $\upsilon_{\mathrm{inp}} = 0.4$, spectral radius $ \rho(\mathbf{W}_{\mathrm{res}}) = 0.2 $, bias scaling $\upsilon_{\mathrm{bias}}  = 0.2$.
The key parameters $\alpha$, $\upsilon_{\mathrm{inp}}$, and  $\rho(\mathbf{W}_{\mathrm{res}})$  were found by empirical evaluations and by a grid search process (see Fig.~\ref{fig:auc_params}).
We fix the seed for the pseudo-random generator (for fixing the randomly generated matrices) and find an appropriate value for the regularization parameter that maximizes performance in a validation set: $ \lambda = 0.001 $. 
Note that the reservoir size is not a parameter to be optimized since with regularized training the performance increases asymptotically with reservoir size \cite{Antonelo2016}. Nonetheless, reservoirs with 100 neurons can already provide acceptable performance not far from what you would get with 600 neurons.
% 100 neurons, TEST AUC 0.976543274833 
% 600 neurons, TEST AUC 0.987438121335 

%
Fig.~\ref{fig:ensaios} shows test data related to two compressor performance tests. 
The first two top plots in Fig.~\ref{fig:ensaio1} correspond to the refrigeration capacity and the shell temperature, whereas the next two below show the desired and predicted outputs (left) and the suction pressure and pressure setpoint indicator (right). Thus, there are 4 input variables and 1 output variable. When the cooling capacity enters the steady-state region (top), the desired output switches from zero to one (bottom). Also, the analog network output (thick grey line), given by the affine operation in (\ref{eq:output}) before applying the nonlinearity $g(.)$ can be seen together with the output thresholded at $0$ (grey line), and a second output thresholded at $0.639$ (thick black line). Depending on the chosen threshold, we get earlier or later predictions that are usually associated to refrigeration capacities that are more distant or closer to the steady-state $cap_f$, respectively. Thus, obviously there is a tradeoff between how soon your prediction happens and how close this prediction is to the steady-state.
The dashed vertical grey stick in the bottom of this plot (around time 31 minutes) indicates the expected mean time that a performance test in the training dataset enters the steady-state region. Thus, on average, the RC model should overperform this fixed inference of the steady state (at $t=31$ minutes) assuming that there is predictable information in the flow of input variables to infer the entrance in the steady-state. If this does not happen, then we could state that the model learned nothing from data or that the data is not enough informative.

Another example can be seen in Fig.~\ref{fig:ensaio2}. The first left plot shows a different behavior in the refrigeration capacity signal, that stabilizes in a different point compared to the previous experiment. The suction pressure has also abnormalities which makes the pressure setpoint indicator become high (1) at around $t=43$, then down (0), and then high again (1). This has a direct effect on the inference of the RC model as seen in the left plot of the second row: the output thresholded at $0$ goes high (+1) temporarily before going down (-1) and up (+1) again. The output thresholded at $0.639$ is less sensitive to earlier predictions as noted graphically.

\begin{figure} [htb!]
 \centering  
 \includegraphics[width=6.8cm]{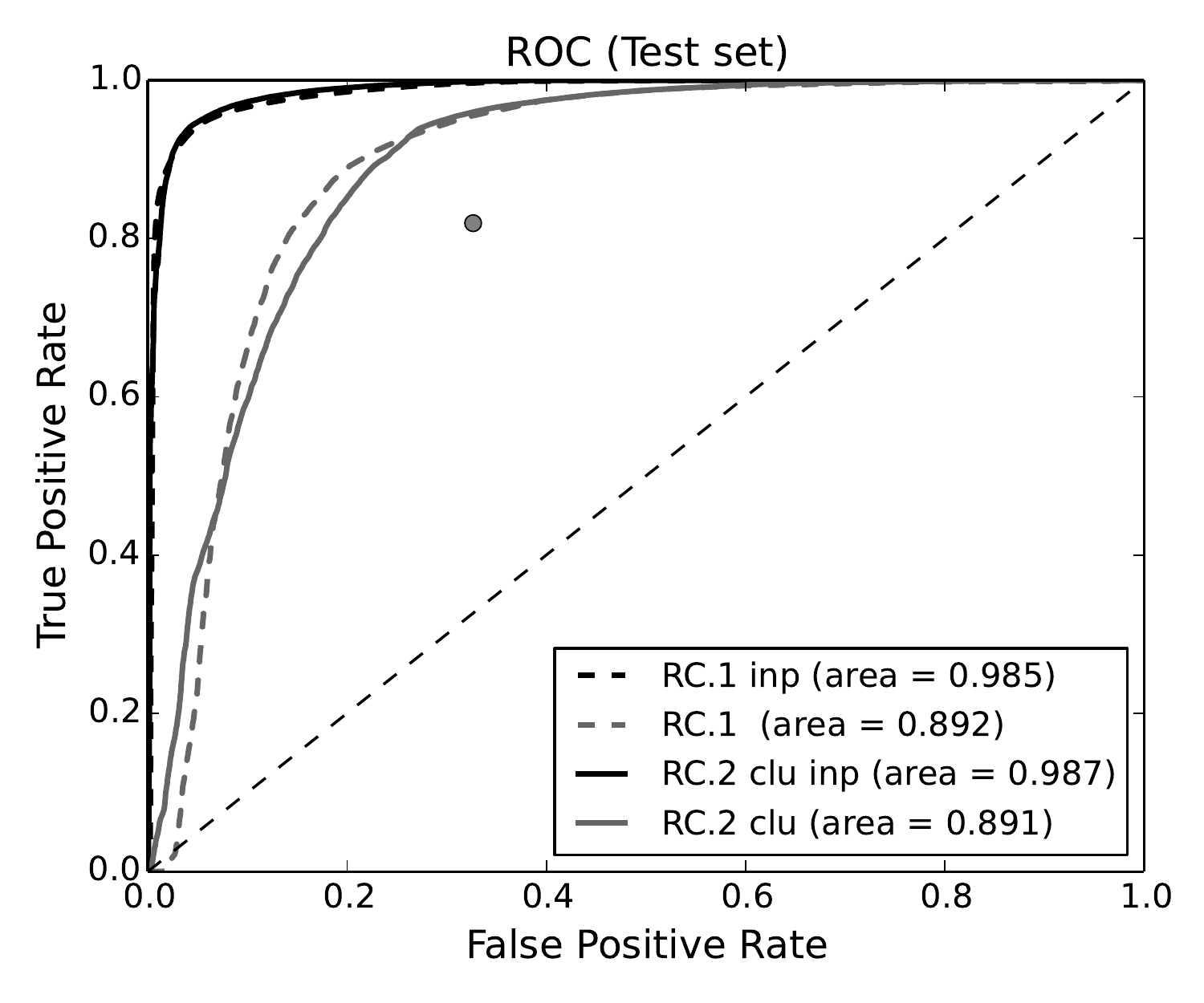}  % 6.1cm
 \caption{ROC curves on test data for $RC.1$, ${RC.1.inp}$ (standard RC model), $RC.2_{clu}$, $RC.2_{clu}.inp$ (proposed RC model). See text for more information.  }
 \label{fig:roc}
\end{figure}

In Fig.~\ref{fig:roc}, we compared the performance of different models in terms of the Receiver Operating Characteristic (ROC) curve and the Area under the ROC curve (or AUC) \cite{hanley1982AUC}. The ROC curve allows us to visually compare models in terms of the True Positive Rate (TPR=TP/(TP + FN)) and of the False Positive Rate (FPR = FP/(TN+FP))\footnote{TP and TN refer to the number of true positives and true negatives (samples correctly predicted in classes +1 and -1), respectively, while FP and FN are the number of false positives and false negatives.} \cite{bradley1997AUC}. 
For each point in the ROC curve, there is a corresponding decision threshold applied to the model output. In general, we want to maximize the TPR while minimizing the FPR and, thus, decision thresholds with corresponding points closer to the top left corner of the ROC plot are usually preferred.
The AUC metric is a very suitable evaluation metric in machine learning \cite{bradley1997AUC}, having desirable properties when compared to the overall accuracy: is threshold independent, and takes into account all the contingencies (TP,FP,TN,FN) when compared to the F1-measure \cite{powers2011evaluation}. 
In Fig.~\ref{fig:roc}, the models called $RC.1$ and $RC.2_{clu}$ correspond to a standard RC network (dashed lines) and the proposed RC model with subspace projection and clustering (from Section~\ref{sec:subspace}) (solid lines), respectively. The variations $RC.1.inp$ and $RC.2_{clu}.inp$ are models that consider the pressure setpoint indicator as input variable to the model (totaling 4 input variables), whereas the others do not (thus, having only 3 input variables). The effect of this setpoint indicator on the performance improvement is clear as the ROC curves (black lines) are drawn closer to the top left corner when compared to the curves without this indicator variable (grey lines).
Additionally, for comparison, we compute performance measures on a \emph{reference naive} model which always infers the steady state at $t=31$ minutes (that is the average expected time that an experiment takes to enter the steady-state). This reference model is represented by a point in Fig.~\ref{fig:roc}, which is the furthest in relation to the top left corner of the graph, showing that all RC models are better predictors than this reference model, i.e., there is informative, predictive value in the stream of measurements that is being captured by the RC models.
While reservoirs with 600 neurons obtain an AUC performance of 0.987, smaller reservoirs present only slightly less performance: reservoirs with 100 neurons and 50 neurons yield AUC of 0.976 and 0.957, respectively.

%%%%%%%%%%%%%%%%%%%%%%%%%%%%%%%%%%%%
%%%%%%%%%%%%%%%%%%%%%%%%%%%%%%%%%%%%

\begin{figure} [htb!]
 \centering  
 \includegraphics[width=7.2cm]{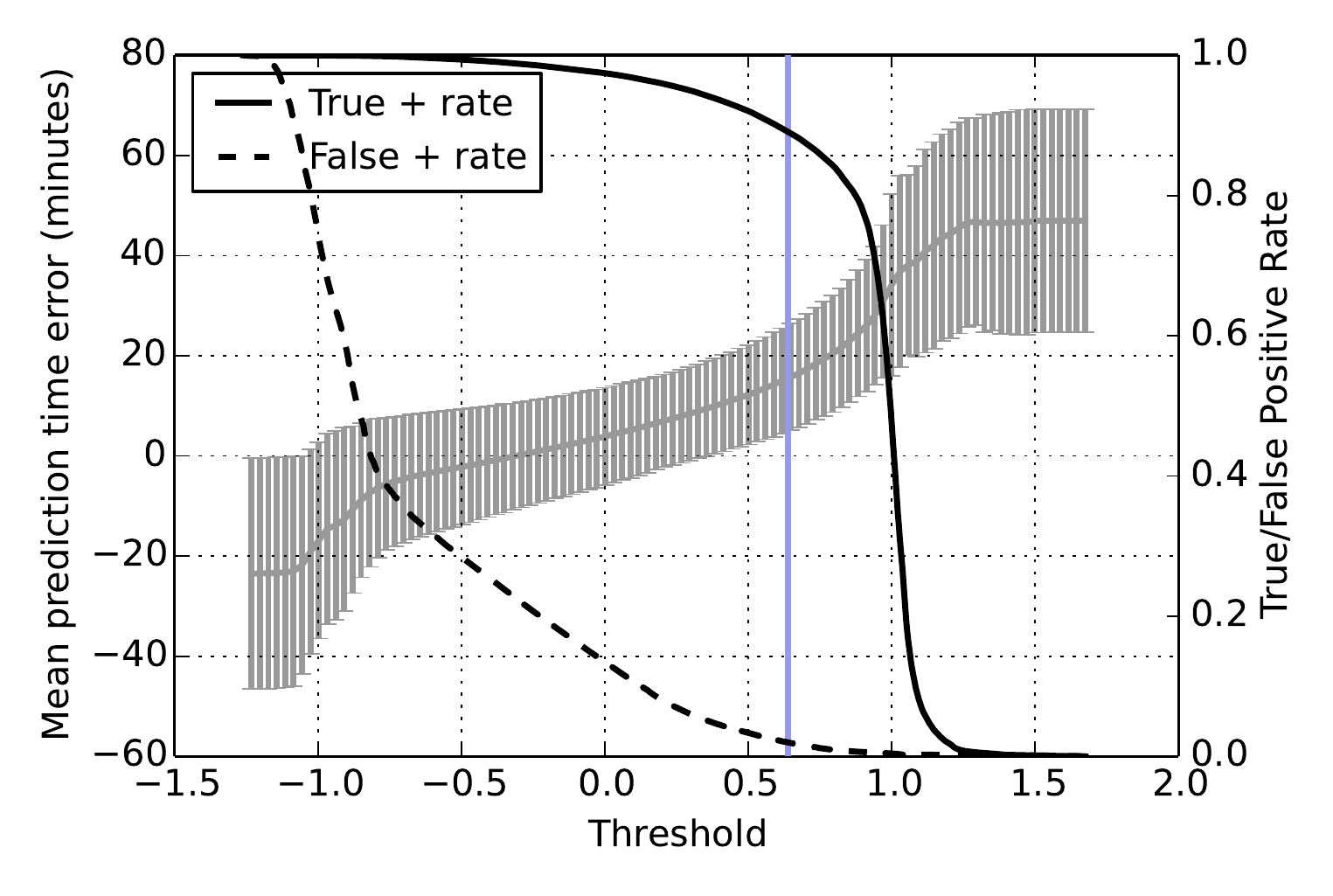}  % 6.1cm
 \caption{True positive (solid black curve) and false positive (dashed black curve) rates (right axis) for different thresholds together with the mean prediction time error with error bars (left axis and grey lines) using $RC.1.inp$ model on test data. The vertical (blue) line marks the threshold (0.639) in which the false positive rate is 0.02 ($2\%$). }
 \label{fig:truepos}
\end{figure}

Another way to view the information from the ROC curve is to plot the true positive (True +) rate and the false positive (False +) rate versus the threshold applied to the analog network output (Fig.~\ref{fig:truepos}). In this figure, the mean prediction time error $t_p - t_d$ is also drawn with error bars denoting the standard deviation. Here, $t_d$ corresponds to the desired (or target) switch time in a episode in which the refrigeration capacity enters the steady state region ($\pm 2\%$ margin), and $t_p$ is the predicted switch time from 0 to 1 given by the thresholded output of the network. We can observe that as the threshold gets higher, the network postpones its prediction on steady-state, making the mean predicted time error also increase. Nevertheless, in practice, 
the requirements of the industrial compressor test prefer models that have a positive average prediction time error since a too early prediction is much more likely to be outside the $\pm 2\%$ margin.
For instance, we could decide that our model should not have a expected false positive rate higher than 0.02. Then, the corresponding threshold that we should apply to the network output is $0.639$, represented by the blue vertical line in Fig.~\ref{fig:truepos}. 
Table~\ref{table:results_fp2} presents some results considering the threshold 0.639 for $RC.1.inp$ and 0.64 for $RC.2_{club}.inp$, both giving a false positive rate of 2\%, a true positive rate of $89\%$ and $88.1\%$, and expected time savings of 86.6 minutes and 85.43 minutes (per performance test), respectively.
Additionally, the corresponding mean prediction time error $\mu_{terr}$ ($terr = t_p - t_d$) are 15.56 minutes for $RC1.inp$  and 16.77 minutes for $RC2_{clu}.inp$, where $\sigma_{terr}$ is the standard deviation.
From this table, we can observe how much time can be saved on average in a performance test and also the mean prediction time error if we use the RC models for steady-state inference and considering that we are willing to accept a false positive rate of only 2\%. From the evaluation metrics shown in this table, we see that both models achieve comparable performance. 
%%%%%%%%%%%%%%%%%%%%%%%%%%%%%%%%%%%%
%%%%%%%%%%%%%%%%%%%%%%%%%%%%%%%%%%%%
%%%%%%%%%%%%%%%%%%%%%%%%%%%%%%%%%%%%
\begin{table}[htp]
\caption{Performance results on test set - False positive rate: 2\% } 
\begin{center}
\begin{tabular}{c|c|c}
\hline
& $RC1.inp$  & $RC2_{clu}.inp$.  \\
\hline
Threshold				      & 0.639   &  0.64   \\
True + rate                            & 89\% & 88.1\%  \\
$\mu_{terr}$ (m)        & 15.56  &   16.77  \\
$\sigma_{terr}$ (m)    & 10.64  & 12.97     \\

Mean time saved (m)                  & 86.6    & 85.43    \\
\hline
\end{tabular}
\end{center}
\label{table:results_fp2}
\end{table}%

On the other hand, if we are willing to accept a higher false positive rate at the threshold equal to zero, we obtain smaller mean prediction time errors and an increase in the mean time saved when the RC models are used. Additionally, setting the threshold of the RC models to zero allows us to compare them against the \emph{Reference} naive model that is built by applying a fixed inference of steady-state always at 31 minutes since the beginning of the episode (which is the expected mean inference time computed from the training dataset).
The statistics of this new setting comparing three models are shown in Table~\ref{table:results_t0}. 
We can clearly observe that $RC1.inp$ and $RC2_{clu}.inp$ are predictive models that have learned informative decisions, presenting zero-one loss classification rate of 4.72\% and 4.26\%, respectively, when compared to the $Reference$ model (zero-one loss rate of 15.24\%). The mean prediction time error $\mu_{terr}$ on test data at threshold 0 are the positive values 3.90 and 1.59 for $RC1.inp$ and $RC2_{clu}.inp$, and -1.90 for the $Reference$ model. Note that the corresponding standard deviation is much smaller for the RC models (9 minutes) than the Reference model (23 minutes).
The percentage of samples where $t_p > t_d$, i.e., the prediction happens after the target desired time ($terr > 0$), is 75\% and 68.47\% for the RC models and 50.74\% for the $Reference$ model. Note that we would like these percentages be as high as possible while still keeping $\mu_{terr}$ as low as possible. 
Another metric relates to how much time we can save assuming that the compressor performance test would stop at the time $t_p$ of steady-state predicted by the models, that is, time saved=$n_e - t_p$. The RC models can save up to 100 minutes on average.
Furthermore, we can see the effect of later predictions on the average time saved per performance test: 98.31 minutes for $RC.1.inp$, and 104.54 minutes for the Reference model. 
Nevertheless, this difference is small and the predictive power of the RC models more than compensates these 5 minutes difference.

\begin{table}[htp]
\caption{Performance results on test set - Threshold: 0 } 
\begin{center}
\begin{tabular}{c|cccc}
\hline
& $Reference$ & $RC.1.inp$  & $RC.2_{clu}.inp$  \\
\hline

0-1 loss                            & 15.24\% & 4.72\% &  4.26\%  \\
AUC                                 & ----   & 0.985  &  0.987    \\
$\mu_{terr}$ (m)        & -1.90  &  3.90  &  1.59   \\
$\sigma_{terr}$ (m)    & 23.06  & 9.75   &  9.12   \\
\hline
Samples where $t_p > t_d$          &  50.74 \%  & 75\% & 68.47\%   \\
Mean time saved (m)         & 104.54 &  98.31 &  100.65  \\
\hline 
\end{tabular}
\end{center}
\label{table:results_t0}
\end{table}%
%%%%%%%%%%%%%%%%%%%%%%%%%%%%%%%%%%%%

We have seen that the standard RC model (\rca) and the proposed one (\rcb) have similar results, with slightly better metrics for the \rcbi~ (e.g., smaller $\mu_{terr}$, 0-1 loss and $\sigma_{terr}$ and higher AUC).
But when we analyze the performance in terms of AUC on a validation dataset for different settings of the spectral radius $ \rho( \mathbf{W}_{\mathrm{res}}) $ and the leak rate $\alpha$ in the reservoir, we get a more insightful perspective on the effect of the self-organized subspace projection on the given task.
Fig.~\ref{fig:auc_params} shows this analysis: \rcbi~has a much smoother performance surface compared to \rcai~while showing high AUC values for most of the configurations of spectral radius and leak rate. On the other hand, \rcai~has AUC values from 0.91 to 0.98 (in comparison, the smallest AUC is about 0.96 in \rcbi). Thus, the proposed method seems to make the modeling task more robust with respect to varying reservoir parameters, yielding some sort of self-organized optimization of the reservoir operation.

%%%%%%%%%%%%%%%%%%%%%%%%%%%%%%%%%%%%
%%%%%%%%%%%%%%%%%%%%%%%%%%%%%%%%%%%%

\begin{figure} [htb!]
 \centering  
 \includegraphics[width=5.8cm]{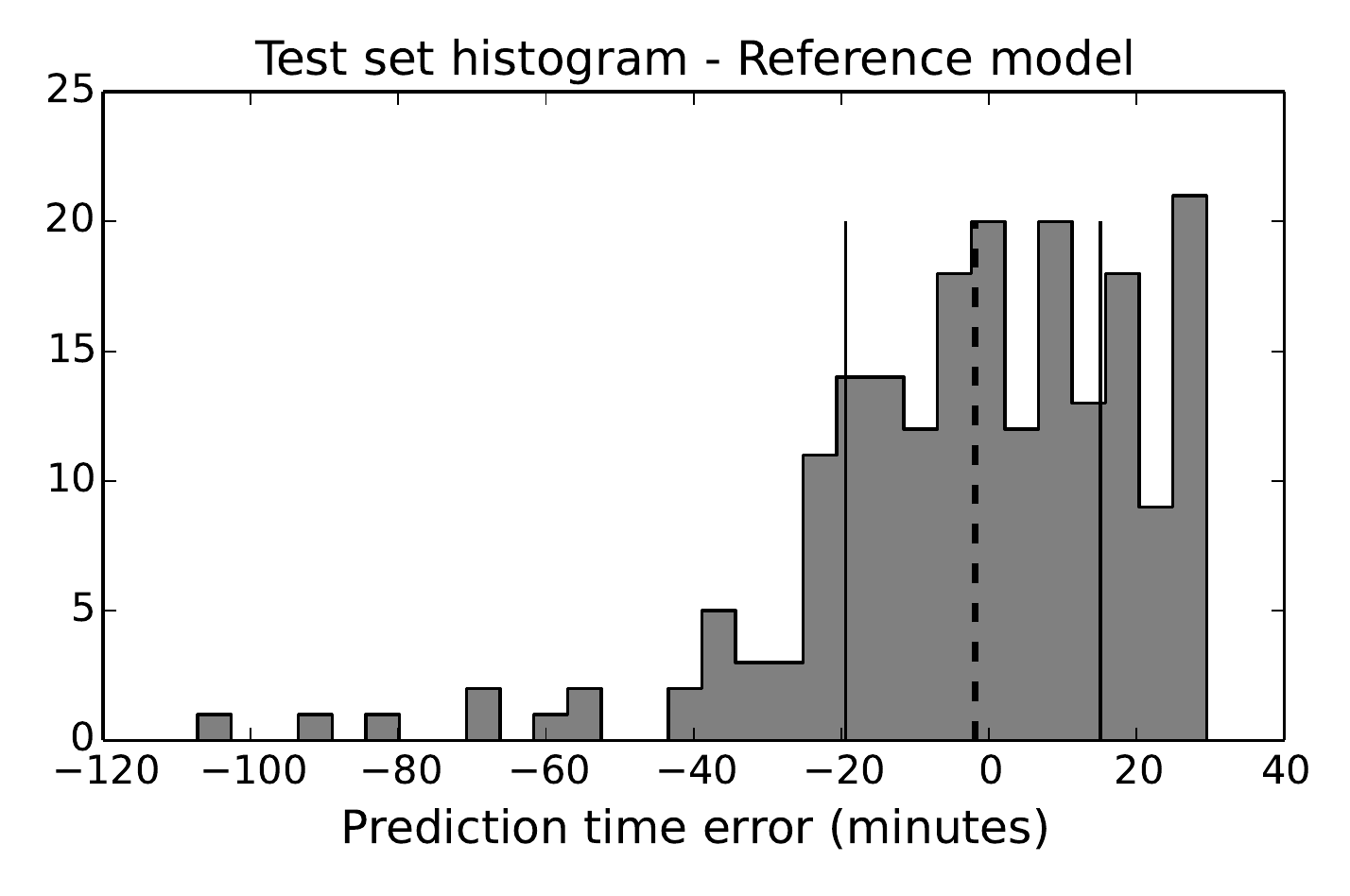}  % 6.1cm
 \includegraphics[width=5.8cm]{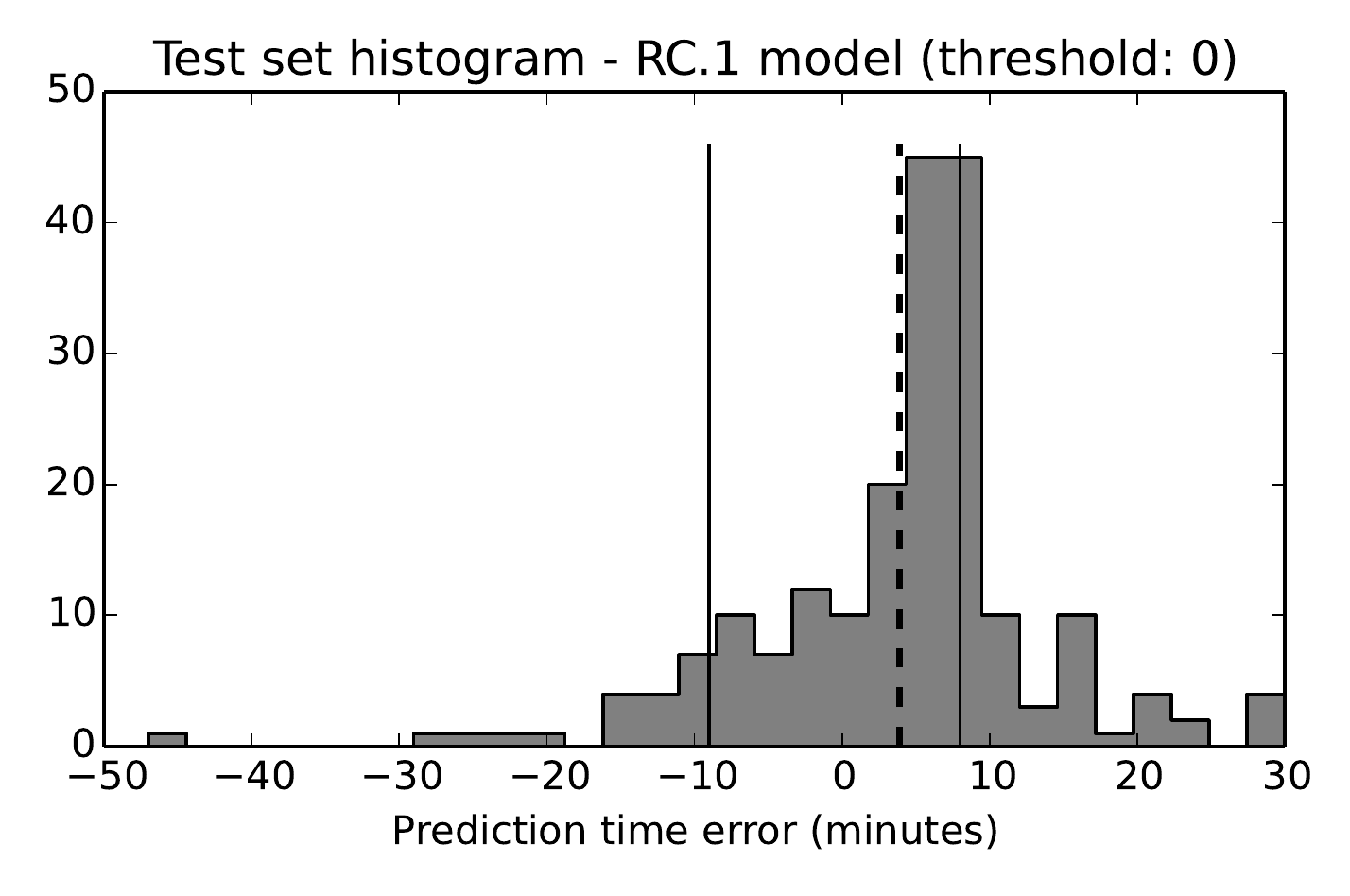}  %
 \includegraphics[width=5.8cm]{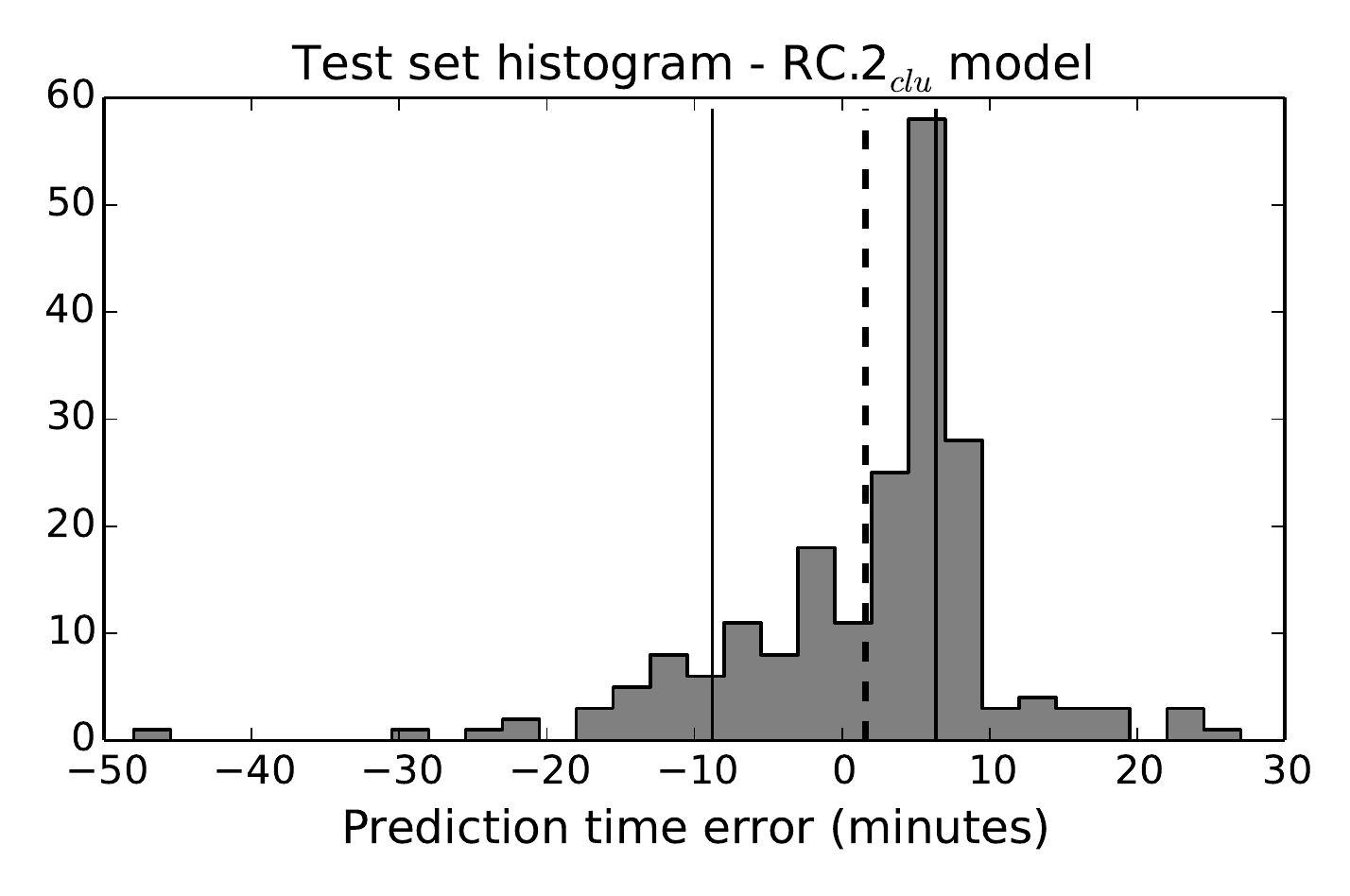}  % 
 \includegraphics[width=5.8cm]{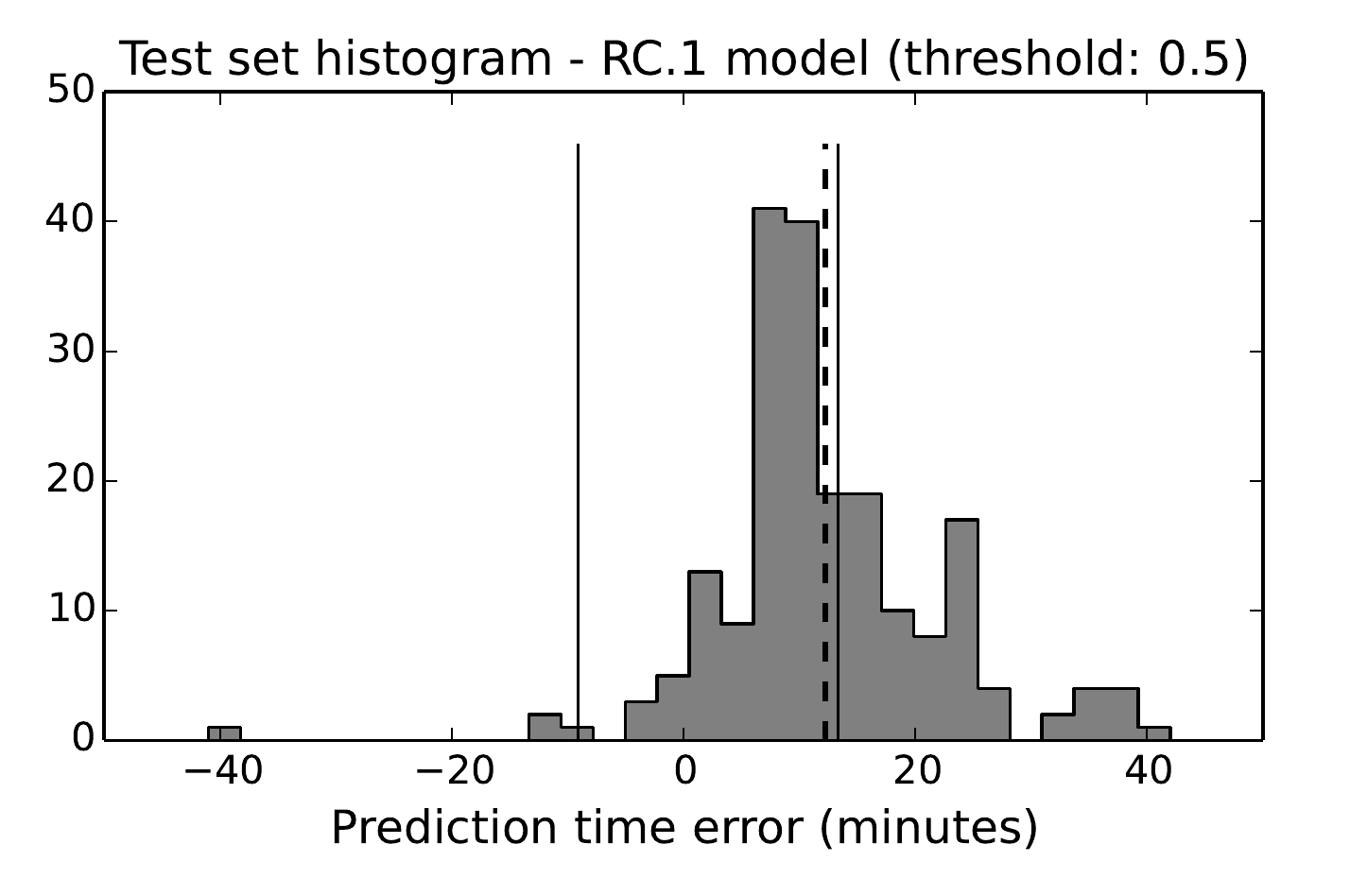}  %
 \caption{Histograms of the prediction time error ($terr=t_p - t_d$) in minutes for different models on the test dataset. The dashed vertical line corresponds to the mean of the error while the solid vertical lines indicate the mean of the error when $terr < 0$ and $terr > 0$, respectively. }
 \label{fig:predtime_err}
\end{figure}

The distribution of the prediction time error $terr$ on the test dataset is shown in Fig.~\ref{fig:predtime_err} for different models and thresholds. All RC models use the pressure setpoint indicator as input variable. Here, we would favor distributions that are mostly centered slightly after $terr=0$ (i.e., $terr > 0$), as given by the RC models.

%%%%%%%%%%%%%%%%%%%%%%%%%%%%%%%%%%%%
%%%%%%%%%%%%%%%%%%%%%%%%%%%%%%%%%%%%

\begin{figure} [htb!]
 \centering  
 \includegraphics[width=5.8cm]{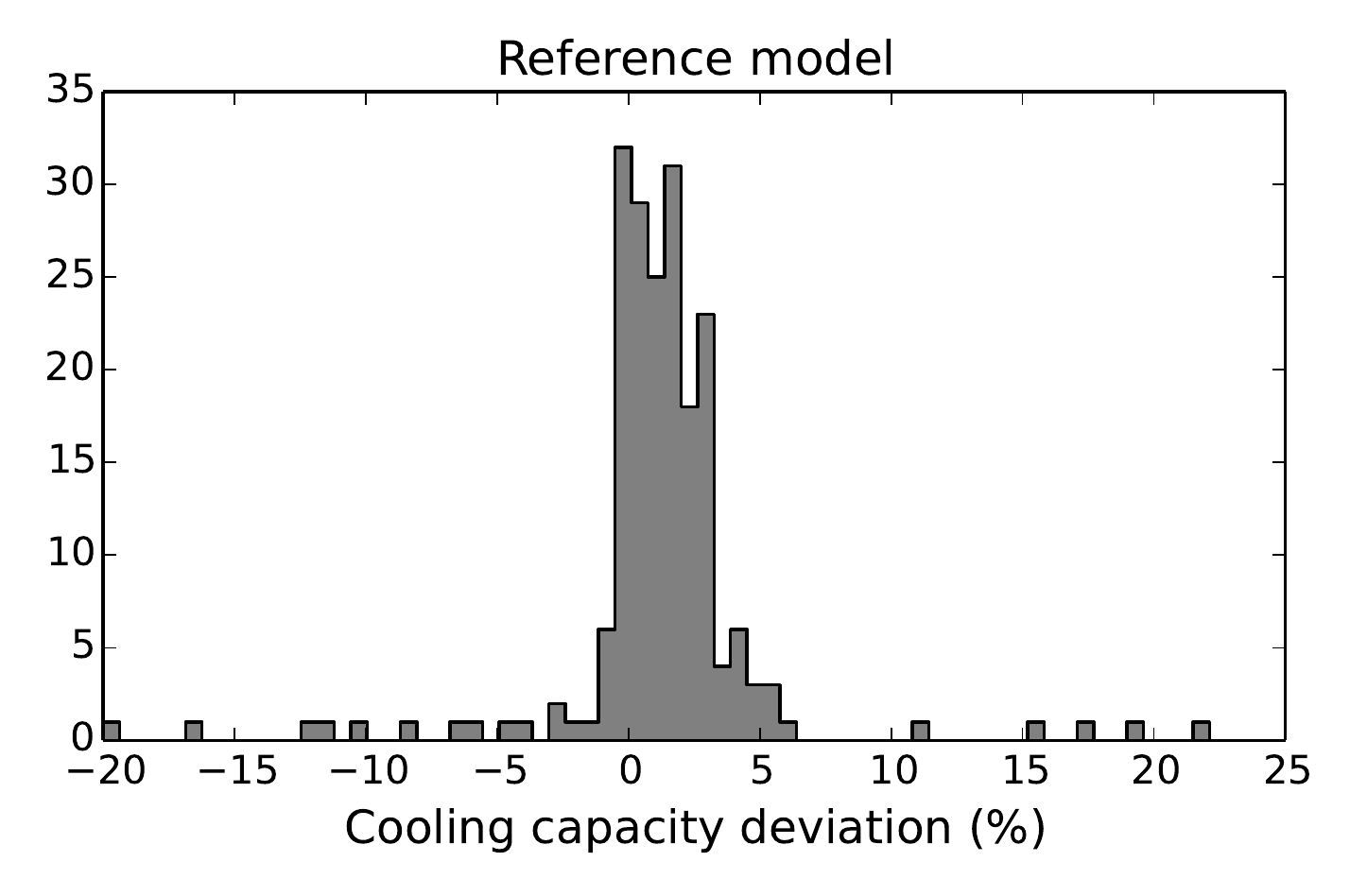}  % 6.1cm
 \includegraphics[width=5.8cm]{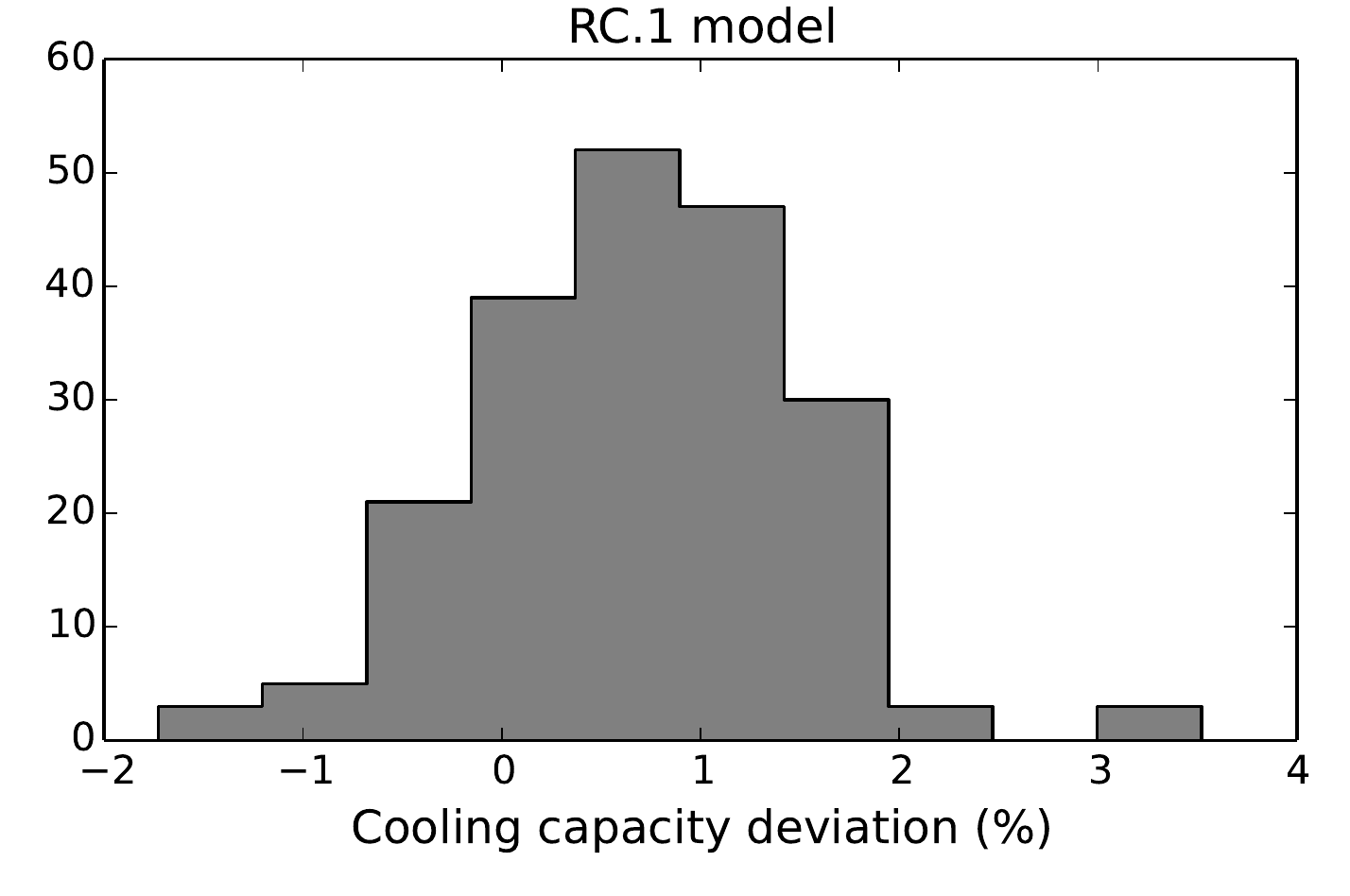}  %  This is closer to steady than RC.2 because, on average, it waits more time to predict (see u_terr, and time saved in table 2)
 \includegraphics[width=5.8cm]{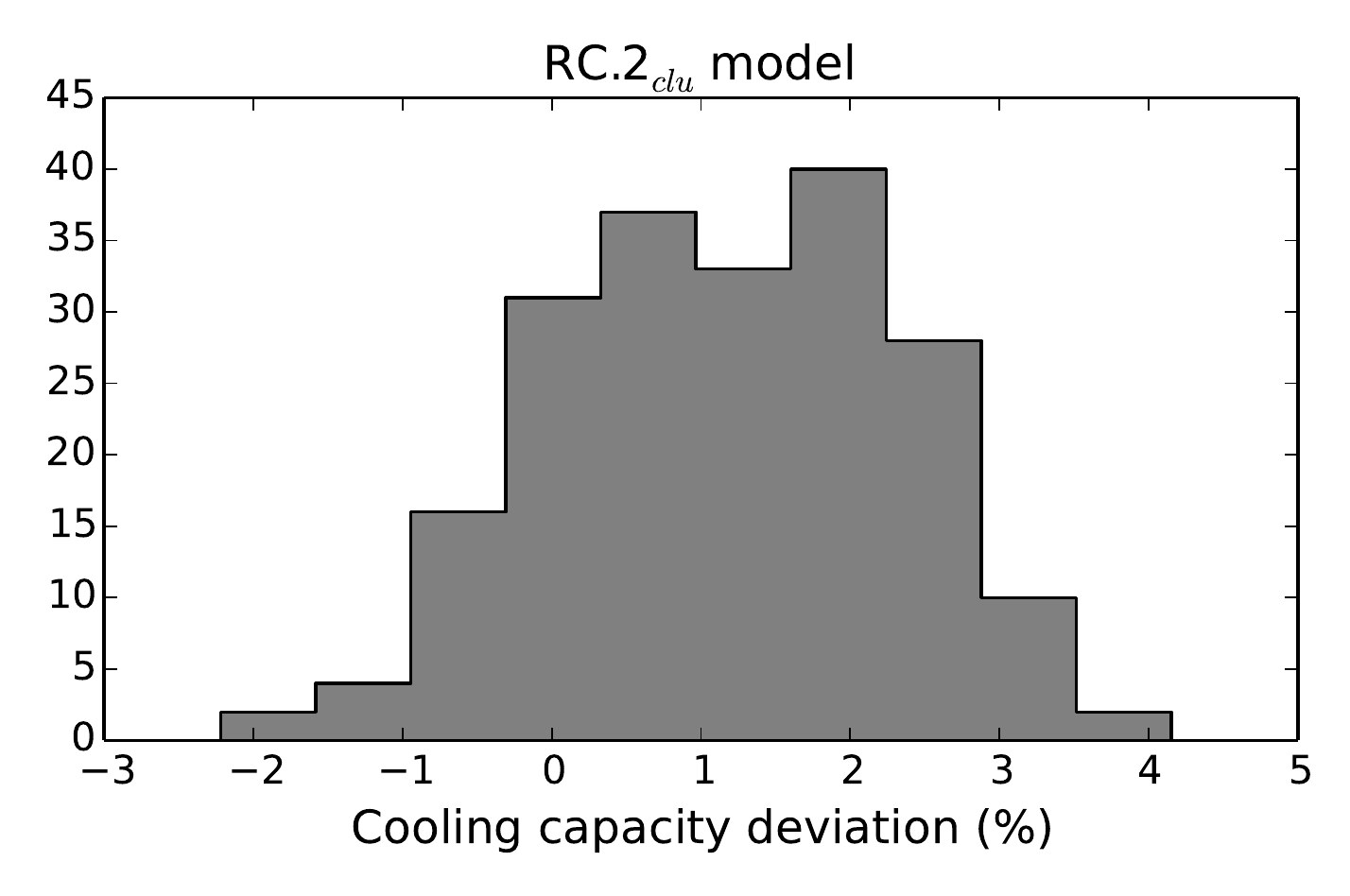}  % 
 \caption{Histograms of the cooling capacity deviation ($(cap(t_p)-cap_f)/cap_f$) for the different models on the test dataset.}
 \label{fig:cap_deviation}
\end{figure}

An additional and important histogram can be made on the cooling capacity deviation ($(cap(t_p)-cap_f)/cap_f$). This allows us to check whether the refrigeration capacity is mostly inside the $\pm 2\%$ margin that characterizes the steady-state condition as defined previously. Fig.~\ref{fig:cap_deviation} shows these distributions obtained from predictions on the test dataset. We can note that in the major part of the compressor tests or episodes, the refrigeration capacity at the time of steady-state entrance detection is inside the $\pm 2\%$ margin. The naive reference model has a much worse distribution, which is understandable as it outputs one always at a fixed time (31 minutes) for all episodes.  

%%%%%%%%%%%%%%%%%%%%%%%%%%%%%%%%%%%%
%%%%%%%%%%%%%%%%%%%%%%%%%%%%%%%%%%%%

\section{Conclusion}
\label{sec:conclusion}
This paper has proposed a self-organized subspace projection method for Reservoir Computing (RC) networks. It builds upon an RC architecture from \cite{Antonelo2011b} with binary inputs for subspace projection applied to robot navigation behaviors. In the current work, the application of the subspace projection originates from an unsupervised learning method (e.g., k-means clustering) which sets the binary input value ($\mathbf{u_2}(t)$) according to the cluster the episode belongs to. Only the initial samples (the first 80 samples or 13.3 minutes) of the cooling capacity in the episode are employed for cluster prediction and corresponding formation of the binary input. Afterwards, the binary input remains fixed until the episode finishes. This actually confines the reservoir operating point to a subspace of trajectories \cite{Antonelo2011b}. 

Both a standard RC model and the proposed method were evaluated on a task of detection of steady-state in performance test of compressors. The data comes from thousands of measurement trials of compressors located in industrial plants from Embraco. The input variables suction pressure, shell temperature and refrigeration capacity were used to detect when the refrigeration capacity enters the steady state region. The results have shown that the flow of measurements provide predictive information for the RC models to stop the compressor test when the cooling capacity is supposed to be entering the 2\% margin of the final cooling capacity (which is unknown beforehand). This means that a significant amount of time can be saved when performing these tests in the industrial plant. 

Future work will tackle to improve the distribution of cooling capacity deviation in order to have a better estimate of the final refrigeration capacity. In the industrial environment, it is worth to also have a model which outputs the reliability of the steady-state detection model so that the prediction of final refrigeration capacity is only used when it is considered reliable.
The method of self-organized subspace projection in RC presented here can be directly generalized into a wider applicable model. It also share similarities to the conceptors model for controlling RNNs from Jaeger \cite{Jaeger2014}.

\section*{Acknowledgements}
The authors thank CAPES and EMBRACO for the supporting fellowship, EMBRACO for providing the industrial datasets composed of the compressor performance tests, and Ahryman Nascimento and Bernardo Schwedersky for their support on the understanding of the industrial datasets.

\section*{References}

\bibliography{biblio}

\end{document}